\documentclass[a4paper,fleqn,usenatbib]{mnras}

\usepackage{newtxtext,newtxmath}

\usepackage[T1]{fontenc}
\usepackage{ae,aecompl}

\usepackage{graphicx}	
\usepackage{amsmath}	
\usepackage{amssymb}	
\usepackage[para,flushleft]{threeparttable}
\usepackage{array,multirow}
\usepackage[labelsep=period,labelfont=bf,font=small]{caption}

\usepackage{natbib,twoopt}
\bibpunct{(}{)}{;}{a}{}{,} 
\makeatletter
\newcommandtwoopt{\citeads}[3][][]{\href{http://adsabs.harvard.edu/abs/#3}%
{\def\hyper@linkstart##1##2{}%
\let\hyper@linkend\@empty\citealp[#1][#2]{#3}}}
\newcommandtwoopt{\citepads}[3][][]{\href{http://adsabs.harvard.edu/abs/#3}%
{\def\hyper@linkstart##1##2{}%
\let\hyper@linkend\@empty\citep[#1][#2]{#3}}}
\newcommandtwoopt{\citetads}[3][][]{\href{http://adsabs.harvard.edu/abs/#3}%
{\def\hyper@linkstart##1##2{}%
\let\hyper@linkend\@empty\citet[#1][#2]{#3}}}
\newcommandtwoopt{\citeyearads}[3][][]%
{\href{http://adsabs.harvard.edu/abs/#3}
{\def\hyper@linkstart##1##2{}%
\let\hyper@linkend\@empty\citeyear[#1][#2]{#3}}}
\makeatother

\title[The beating magnetic heart in Horologium]{Far beyond the Sun: I. The beating magnetic heart in Horologium}

\author[J. D. Alvarado-G\'omez et al.]{Juli\'an D. Alvarado-G\'omez$^{1}$\thanks{E-mail: julian.alvarado-gomez@cfa.harvard.edu},
Gaitee A. J. Hussain$^{2,3}$,
Jeremy J. Drake$^{1}$,
\newauthor
Jean-Fran\c{c}ois Donati$^{4}$,
Jorge Sanz-Forcada$^{5}$,
Beate Stelzer$^{6}$,
Ofer Cohen$^{7}$,
\newauthor
Eliana M. Amazo-G\'omez$^{8,9}$,
Jason H. Grunhut$^{10}$,
Cecilia Garraffo$^{1}$,
\newauthor
Sofia P. Moschou$^{1}$,
James Silvester$^{11}$, and
Mary E. Oksala$^{12,13}$
\\
\\
$^{1}$Harvard-Smithsonian Center for Astrophysics, 60 Garden Street, Cambridge, MA 02138, USA\\
$^{2}$European Southern Observatory, Karl-Schwarzschild-Str. 2, 85748 Garching bei M\"unchen, Germany\\
$^{3}$Institut de Recherche en Astrophysique et Plan\'etologie, Universit\'e de Toulouse, UPS-OMP, F-31400 Toulouse, France\\
$^{4}$CNRS-IRAP, 14, avenue Edouard Belin, F-31400 Toulouse, France\\
$^{5}$Centro de Astrobiolog\'ia (CSIC-INTA), ESAC Campus, Camino Bajo del Castillo, E-28692 Villanueva de la Ca{\~n}ada, Madrid, Spain\\
$^{6}$Eberhard Karls Universit\"{a}t, Institut f\"{u}r Astronomie und Astrophysik, Sand 1, 72076 T\"{u}bingen, Germany\\
$^{7}$University of Massachusetts at Lowell, Department of Physics \& Applied Physics, 600 Suffolk St., Lowell, MA 01854, USA\\
$^{8}$Max-Planck-Institut f\"{u}r Sonnensystemforschung, Justus-von-Liebig-Weg 3, 37077 G\"{o}ttingen, Germany\\
$^{9}$Georg-August Universit\"{a}t G\"{o}ttingen, Institut f\"{u}r Astrophysik, Friedrich-Hund-Platz 1, 37077 G\"{o}ttingen, Germany\\
$^{10}$Dunlap Institute for Astronomy and Astrophysics, University of Toronto, 50 St. George Street, Toronto, ON, M5S 3H4, Canada\\
$^{11}$Department of Astronomy and Space Physics, Uppsala University, SE-751 20 Uppsala, Sweden\\
$^{12}$LESIA, Observatoire de Paris, PSL Research University, CNRS, Sorbonne Universit\'es, UPMC, Univ. Paris 6, Univ. Paris Diderot,
\\\hspace{0.34cm}Sorbonne Paris Cit\'e, 5 place Jules Janssen, F-92195 Meudon, France\\
$^{13}$Department of Physics, California Lutheran University, 60 West Olsen Road \#3700, Thousand Oaks, CA, 91360, USA
}

\date{Accepted 2017 October 5. Received 2017 October 5; in original form 2017 August 2}

\pubyear{2017}

\begin{document}
\label{firstpage}
\pagerange{\pageref{firstpage}--\pageref{lastpage}}
\maketitle

\begin{abstract}
A former member of the Hyades cluster, $\iota$ Horologii ($\iota$ Hor) is a planet-hosting Sun-like star which displays the shortest coronal activity cycle known to date ($P_{\rm cyc}\sim$\,1.6~yr). With an age of $\sim$$625$~Myr, $\iota$ Hor is also the youngest star with a detected activity cycle. The study of its magnetic properties holds the potential to provide fundamental information to understand the origin of cyclic activity and stellar magnetism in late-type stars. In this series of articles, we present the results of a comprehensive project aimed at studying the evolving magnetic field in this star and how this evolution influences its circumstellar environment. This paper summarizes the first stage of this investigation, with results from a long-term observing campaign of $\iota$ Hor using ground-based high-resolution spectropolarimetry. The analysis includes precise measurements of the magnetic activity and radial velocity of the star, and their multiple time-scales of variability. In combination with values reported in the literature, we show that the long-term chromospheric activity evolution of $\iota$ Hor follows a beating pattern, caused by the superposition of two periodic signals of similar amplitude at $P_{1} \simeq 1.97 \pm 0.02$~yr and $P_{2} \simeq 1.41 \pm 0.01$ yr. Additionally, using the most recent parameters for $\iota$ Hor~b in combination with our activity and radial velocity measurements, we find that stellar activity dominates the radial velocity residuals, making the detection of additional planets in this system challenging. Finally, we report here the first measurements of the surface longitudinal magnetic field strength of $\iota$ Hor, which displays varying amplitudes within $\pm4$ G and served to estimate the rotation period of the star ($P_{\rm rot} =~7.70^{+0.18}_{-0.67}$~d).

\end{abstract}



\section{Introduction}\label{sect_1}

Early studies performed by \citetads{1968ApJ...153..221W}, \citetads{1980PASP...92..385V}, \citetads{1995ApJ...438..269B}, and \citetads{1996AJ....111..439H}, showed that chromospheric emission and activity cycles are a common feature among late-type stars. Cyclic variability of the Ca II H (396.84 nm) \& K (393.36 nm) emission cores is observed in about $60$ per cent of main-sequence stars in the solar neighbourhood, including spectral types from early M to F (\citeads{1995ApJ...438..269B}). The cycle periods range from $2.5$ to $25$ years, placing the $\sim$$11$-yr solar activity cycle in the middle of the distribution, and the cycle occurrence rate is related to the mean activity level (and therefore to stellar rotation and age; see \citeads{1972ApJ...171..565S}, \citeads{1984ApJ...279..763N}). Recent efforts are being carried out in order to extend the baseline of observations and increase the sample of identified cycles (e.g. Hall~et~al.~\citeyearads{2007AJ....133..862H}, \citeyearads{2009AJ....138..312H}, \citeads{2012IAUS..286..317M}), and to place the solar chromosphere in the general stellar context (e.g. \citeads{2016ApJ...826L...2M}, \citeads{2016csss.confE...6E}).

The standard picture for the origin of these stellar activity cycles comes from our knowledge of the behaviour of the Sun's magnetism. It is widely accepted that the $11$-yr global activity changes of the Sun appear as a consequence of a $22$-yr magnetic cycle which, among other features, is primarily characterised by a double polarity reversal of the small-\footnote[1]{The terms small- and large- are used here in the stellar context; the former indicates spots/active region size while the latter is used for the star as a whole.} and large-scale fields (\citeads{2012LRSP....9....6M}, \citeads{2015LRSP...12....4H}). The latter are nowadays accessible in the stellar case via spectropolarimetric observations and the technique of Zeeman Doppler Imaging (ZDI, \citeads{1997MNRAS.291..658D}, \citeads{2002A&A...381..736P}, \citeads{2009MNRAS.398..189H}). While large-scale polarity reversals have been detected in a handful of objects (see \citeads{2016IAUFM..29A.360F}), in only one of them solar-like cyclic behaviour has been reported ($61$ Cyg A; see \citeads{2016A&A...594A..29B}). For the majority of targets the reversal time-scale is typically very short ($\sim$$1-2$ yr) and disconnected from the observed variability in the chromospheric activity, showing no resemblance in this regard to the solar case. It is unclear whether these reversals may have any connection with either stellar equivalents of the solar quasi biennial oscillation (see \citeads{2014SSRv..186..359B}, \citeads{2015NatCo...6E6491M}), a different type of magnetism-activity relation in active stars, or an observational bias introduced by the methods used in these comparisons. Efforts are still ongoing in order to connect the properties of the recovered magnetic fields with the characteristics of the stellar activity cycles in these systems (see~\citeads{2016MNRAS.462.4442S}).

Following the solar analogy, a clearer signature of these stellar chromospheric activity cycles is expected from the coronal high-energy emission (e.g. UV/X-rays). Indeed, the amplitude of the solar chromospheric Ca H\,\&\,K cycle is only few per cent of the mean value (see \citeads{2017ApJ...835...25E} and references therein), while in coronal X-rays, particularly in the soft X-ray band (i.e., $1 - 8$ \AA), cycle amplitudes between $0.7 - 2.0$ dex have been reported (\citeads{1994SoPh..152...53A}, \citeads{2001ApJ...560..499O}, \citeads{2003ApJ...593..534J}). Unfortunately, even in the optimistic scenario of solar-like cycle amplitudes, the detection of stellar X-ray cycles is greatly compromised by the difficulty in maintaining homogeneous (i.e. with the same instrument) and continuous monitoring over cycle time-scales. The increased coverage achieved by combining datasets from different instruments involves non-trivial cross-calibration procedures, which are not only cumbersome but also may introduce dominant spurious signals to the data (e.g. the case of $\alpha$~Cen A, see \citeads{2008ApJ...678L.121A}). 

For these reasons is not surprising that X-ray activity cycles have been identified only in $5$ late-type stars\footnote[2]{Possibly $6$ if the tentative $\sim$$19$-year cycle of $\alpha$ Cen A is confirmed (\citeads{2014AJ....147...59A}).}. Three members of this group, namely HD~81809 (G2V + G9V, $P_{\rm cyc} = 8.2$ yr; Favata~et~al.~\citeyearads{2004A&A...418L..13F}, \citeyearads{2008A&A...490.1121F}), $61$~Cyg~A (K5V, $P_{\rm cyc} = 7.3$ yr; \citeads{2006A&A...460..261H}, \citeads{2012A&A...543A..84R}), and $\alpha$~Cen~B (K1V, $P_{\rm cyc} = 8.1$ yr; Ayres~\citeyearads{2009ApJ...696.1931A}, \citeyearads{2014AJ....147...59A}, \citeads{2010ApJ...722..343D}, \citeads{2012A&A...543A..84R}), belong to binary systems (where non cycle-related mechanisms could be involved in the X-ray variability). Recently, \citetads{2017MNRAS.464.3281W} reported an X-ray activity cycle of $P_{\rm cyc} = 7.1$ yr for the fully convective M-dwarf Proxima Cen (M5.5V), coincident with the long-term photometric variability of the star (\citeads{2016A&A...595A..12S}). The remaining star in this sample is $\iota$ Horologii (HD~17051~=~HR~810, G0V), which not only displays the shortest X-ray cycle known to date ($P_{\rm cyc}~=~1.6$~yr, \citeads{2013A&A...553L...6S}), but also is the youngest star with a detected cycle ($\sim$$625$~Myr, see~Table~\ref{tab_1}). The $1.6$-yr activity cycle of $\iota$ Hor was first identified by \citetads{2010ApJ...723L.213M} in chromospheric emission and, using additional Ca H\,\&\,K data, \citetads{2017MNRAS.464.4299F} recently suggested the presence of a secondary $\sim$$4.5$-yr periodicity. As very young stars show erratic and non-cyclic activity (see \citeads{1995ApJ...438..269B}, \citeads{1999ApJ...524..295S}, \citeads{2015ARep...59..726K}, \citeads{2016A&A...590A.133O}), the magnetic and activity cycles of $\iota$ Hor could be representative of the onset of cycles during the life of a Sun-like star and provide critical information to understand the generation and evolution of magnetism in late-type stars. 

In this context, this paper is the first in a series of articles containing the results of the \textit{``Far beyond the Sun''} campaign: an observational-numerical study aimed at characterising the magnetic cycle of $\iota$ Hor using high-resolution spectropolarimetry, and determining (from data-driven simulations), how this evolving field influences the corona and wind environment around the star. Here we introduce the observing programme and focus on the different time-scales of variability of the activity indicators, the radial velocity of the star, and the surface-averaged longitudinal magnetic field. The astrophysical properties of $\iota$ Hor are presented in Sect. \ref{sect_2}. Section \ref{sect_3} contains a description of the observations and data processing. A description of the different measurements and their results is provided in Sect. \ref{sect_4}. We analyse and discuss our findings in Sect. \ref{sect_5}, and summarise our work in Sect. \ref{sect_6}.

\section{Astrophysical properties of $\iota$ Hor}\label{sect_2}

The $11^{\rm th}$ brightest star of the Horologium constellation (the pendulum clock), $\iota$ Hor ($V_{\rm mag} = 5.4$) is a young Sun-like star located at $17.17 \pm 0.06$ pc from the Sun (\citeads{2007A&A...474..653V}). Despite its location in the southern hemisphere [$\alpha$~(J2000):~$02^{\rm h}\,42^{\rm m} \,33.47^{\rm s}$, $\delta$ (J2000): $-50^\circ \,48\arcmin\,01.10\arcsec$], strong evidence suggests that the star was formed in the Hyades cluster, currently sharing the same kinematic properties (i.e.~part of the Hyades stream; \citeads{2001MNRAS.328...45M}, \citeads{2004A&A...418..989N}), and having the same metallicity, helium abundance, and age (\citeads{2007A&A...463..657L}, \citeads{2008A&A...482L...5V}). 

The star is also known to host a $\sim$$2.5$ Jupiter-mass planet at approximately $1$ AU. The exoplanet was discovered by \citetads{2000A&A...353L..33K}, and shortly after confirmed independently by \citetads{2001A&A...375..205N} and \citetads{2001ApJ...555..410B}. The most recent orbital parameters of this system ($M_{\rm p}\sin(i) = 2.48 \pm 0.08$ M$_{\rm J}$, $P_{\rm orb} = 307.2 \pm 0.3$ days, $e = 0.18 \pm 0.03$, $a = 0.96 \pm 0.05$ AU), were obtained by \citetads{2013A&A...552A..78Z} using a compilation of radial velocities (RVs) from all previous studies (including additional data presented by \citeads{2006ApJ...646..505B}), together with new RV measurements from ESO's CES and HARPS spectrographs. The combined solution reproduces the $\sim$$15$-yr baseline of RV measurements, although relatively high scatter is obtained in the orbit residuals (with an rms of $14.5$ m s$^{-1}$). \citetads{2013A&A...552A..78Z} acknowledge the presence of an activity cycle of $\iota$ Hor, which shows a roughly $2$:$1$ relation with the orbital period of the planet. Nevertheless they conclude that the activity cycle is not responsible for the observed RV signal, although it can certainly be the dominant cause for the scatter in the residuals, particularly, since the activity-filtering analysis of \citetads{2011A&A...528A...4B} ruled out companions with orbital periods shorter than seven days.      
 
\begin{table}
\caption{Fundamental properties of $\iota$ Hor.}\label{tab_1}      
\begin{threeparttable}
\centering                          
\begin{tabular}{l c c}        
\hline\hline                
Parameter & Value & Reference \\
\hline
Spectral Type & F8V\,--\,G0V & \protect{\citeads{2010MNRAS.405.1907B}}\\ 
$T_{\rm eff}$ [K] & $6080 \pm 80$ & \protect{\citeads{2010MNRAS.405.1907B}} \\
$\log(g)$ & $4.399 \pm 0.022$ & \protect{\citeads{2010MNRAS.405.1907B}} \\
$R_{*}$ [R$_{\odot}$] & $1.16 \pm 0.04$ & \protect{\citeads{2010MNRAS.405.1907B}} \\
$M_{*}$ [M$_{\odot}$] & $1.23 \pm 0.12$ & \protect{\citeads{2010MNRAS.405.1907B}} \\
$v\sin i$ [km s$^{-1}$] & $6.0 \pm 0.5$  & This work \\
$\left<v_{\rm R}\right>$ [km s$^{-1}$]$^a$ & $16.943 \pm 0.002$ & This work \\ 
$P_{\rm rot}$ [days] & $7.70^{+0.18}_{-0.67}$ & This work \\
$\left<\log(L_{\rm X})\right>$ & $28.78 \pm 0.08$ & \protect{\citeads{2013A&A...553L...6S}}\\
Age [Myr]$^b$ & $\sim625$ & \protect{\citeads{2008A&A...482L...5V}}\\
\hline                                   
\end{tabular}
\begin{tablenotes}
{\small $^{(a)}$ Average value from the multi-epoch HARPSpol observations (see Sect. \ref{sect_3}).}
{\small $^{(b)}$ Age derived from HARPS asteroseismology. This value falls in between the estimates from gyrochronology ($\sim$$740$~Myr, \citeads{2007ApJ...669.1167B}), and from the level of X-ray/EUV emission ($\sim$$500$~Myr, \citeads{2011A&A...532A...6S}).}
\end{tablenotes}
\end{threeparttable}
\end{table} 
  
Table \ref{tab_1} contains a summary of the main properties of $\iota$ Hor and their corresponding references. Using the fundamental parameters provided by \citetads{2010MNRAS.405.1907B}, and applying a standard spectral synthesis on a set of iron Fe~{\sc I} and Fe~{\sc II} lines under Kurucz ATLAS9 model atmospheres in LTE, we obtained a projected rotational velocity of $6.0 \pm 0.5$ km s$^{-1}$. This result is consistent with the value reported by \citetads{2005ApJS..159..141V} derived from a similar analysis based on UCLES@AAT spectra (\citeads{1990SPIE.1235..562D}). As discussed in more detail in the next section, we measure the radial velocity on each night and calculate the long-term average $\left<v_{\rm R}\right>$ listed in Table~\ref{tab_1}. Likewise, our analysis indicates a rotation period of $P_{\rm rot} = 7.70^{+0.18}_{-0.67}$ d, roughly consistent with previous reports from \citetads{1997MNRAS.284..803S} and \citetads{2010ApJ...723L.213M}, estimated using different methods (see~Sect.~\ref{sec_rotation}).

\section{Observations and Data Processing}\label{sect_3}

\subsection{Spectropolarimetric data}

\subsubsection{Observing strategy}\label{sec_strategy}

We began the monitoring of the magnetic cycle of $\iota$ Hor in October $2015$, using the spectropolarimetric mode of the HARPS spectrograph (HARPSpol) attached at the ESO $3.6$m telescope located at the La Silla Observatory in Chile (\citeads{2003Msngr.114...20M}, \citeads{2011Msngr.143....7P}). 

Apart from seasonal visibility, two important elements determined the observing strategy for this programme. The first one considered the coverage needed in order to resolve the expected activity cycle evolution, taking the results from the ongoing X-ray monitoring of the star as reference (\citeads{2016csss.confE.112S}). On the other hand, for each observing epoch we required sufficient (rotational) phase sampling to guarantee the successful retrieval of the magnetic field distributions using ZDI\footnote[3]{The recovered ZDI large-scale magnetic field maps for each epoch will be presented in the second paper of this study.} (including possible weather losses). Given the moderate activity levels of $\iota$ Hor (i.e., Ca H\,\&\,K S-index between $0.23-0.28$), and the large amount of telescope time required for a single full Stokes ZDI inversion (even for stars with stronger magnetic fields; see \citeads{2015ApJ...805..169R}), only circular polarization (Stokes V) was considered. 

The exposure time requirements were based on our previous HARPSpol experience mapping magnetic fields of stars with similar activity levels and spectral type (e.g., \citeads{2015A&A...582A..38A}, \citeads{2016A&A...585A..77H}). To achieve the required S/N ($\sim$$400-500$@550 nm), and to prevent any possible saturation of the detector (taking into account the visual brightness of the star; see Sect. \ref{sect_2}), the acquisition of two (or three) Stokes V exposures consecutively during the same night was planned, with a total integration time of $\sim$\,1 hour per night\footnote[4]{Owing to bad weather conditions, on four separate nights only a single Stokes V exposure was retrieved (see Table \ref{table_2}). Likewise, rapidly changing weather during the night of Sep. 13 2016 (BJD:~2457646.77036) required the acquisition of four Stokes V exposures to secure the required S/N level.}. In this way, nine epochs have been secured over a $\sim$$1.4$ yr baseline (between Oct. $2015$ and Feb. $2017$). One epoch consists of roughly two weeks of almost consecutive nights, each one of these composed of $1$$\,-\,$$3$ high S/N merged Stokes V exposures (effectively $4-12$ spectra per night) which, as described below, also yield extremely high S/N unpolarized spectra (Stokes~I). Details for the individual nights can be found in the journal of observations (columns $1-5$, Table~\ref{table_2}).

\subsubsection{Level 1 processing}\label{lv1}

The retrieval of the circularly polarized spectra is performed using the ratio method (see \citeads{1997MNRAS.291..658D}, \citeads{2009PASP..121..993B}). Four sub-exposures, each one consisting of two orthogonal polarization states (carried separately by individual fibers to the spectrograph), are divided coherently to produce a single Stokes V spectrum. By considering a ratio, an effective first order removal of spurious signals and systematic errors is automatically performed. This is complemented with the aid of the so-called null polarization spectrum (denoted by N), generated from the ratio of destructive (incoherent) polarization states. By construction, the N spectrum should remain at the zero level at all times if only polarization from the astrophysical object is considered (deviations from zero are indicative of spurious signals). The Stokes I spectra are generated by simply co-adding all the sub-exposures together, which typically leads to much larger S/N than in standard spectroscopic stellar observations.

The data reduction process was carried out using the automatic {\sc libre-esprit} package (see \citeads{1997MNRAS.291..658D}), which has been recently modified to handle the extraction of HARPSpol observations, preserving the standard RV precision of the instrument (cf. \citeads{2016A&A...585A..77H}, \citeads{2016MNRAS.461.1465H}). The spectra are obtained following an optimal extraction scheme, after bias, flat-field, and cosmic ray corrections are applied. Two different sets of ThAr arc spectra are used to compute the wavelength solution and the corresponding barycentric corrections for each night. The latter are obtained from the JPL Horizons ephemeris database\footnote[5]{\url{https://ssd.jpl.nasa.gov}}, using the information on the stellar exposures and computing the velocity of the observer (and the corresponding time shift) with respect to the barycenter of the solar system. The pipeline applies a raw automatic continuum normalization over the entire spectral range ($378 - 691$ nm), yielding a typical ten per cent error of the continuum level. This was drastically improved with the aid of two additional procedures. First, a refined continuum shape and normalization were determined using the automatic spline fitting algorithm implemented in the \textit{iSpec} package (\citeads{2014A&A...569A.111B}), with the recommended settings for the HARPS spectrograph (one cubic-spline per every nm, $R \sim 115000$). A second re-normalization procedure was then applied by visually inspecting each spectrum over $15$ nm windows, and fitting an additional cubic-spline slowly-varying envelope to the entire wavelength range. In this way, we achieved a typical error of roughly one percent for the continuum determination in the final reduced spectra.

\subsubsection{Level 2 processing}\label{lv2}

\noindent The second step in the data processing corresponds to the extraction of the Zeeman signatures from the polarized spectra. As discussed in detail by \citetads{2009ARA&A..47..333D}, the Zeeman components induced by surface magnetic fields in Sun-like stars such as $\iota$~Hor, are not detectable in individual spectral lines with current instrumentation. The limitation is circumvented with the aid of a multi-line cross-correlation technique known as Least-Squares Deconvolution (LSD, see \citeads{1997MNRAS.291..658D}, \citeads{2010A&A...524A...5K}). This technique adds coherently the polarimetric signature from a large number of spectral lines contained in the echelle spectra, and synthesizes extremely high S/N average line profiles (denoted as LSD profiles) for all the relevant polarimetric quantities in each observation (i.e., Stokes I, Stokes V, and N spectra). 

A photospheric model of the target star (line mask), containing information regarding the rest wavelengths, Land\'e factors, and depths of all the atomic lines within the observed wavelength range, is required to perform the LSD analysis. This was generated using the Vienna Atomic Line Database (VALD3\footnote[6]{\url{http://vald.astro.uu.se}}, \citeads{2000BaltA...9..590K}, \citeads{2015PhyS...90e4005R}), assuming a detection threshold of $0.05$ (in normalized units), a microturbulence of $1.04$ km s$^{-1}$ (\citeads{2010MNRAS.405.1907B}), together with the $T_{\rm eff}$ and $\log(g)$ values listed in Table~\ref{tab_1}. This line mask was optimized following the methodology presented in \citetads{2015A&A...582A..38A}, to closely match all the individual line depths in the observed spectrum. The final mask employed for the extraction of the LSD profiles of $\iota$ Hor contains $8834$ spectral lines in total. The profiles were extracted for the velocity space between $-5.0$ km s$^{-1}$ and $45.4$ km s$^{-1}$, with a velocity step of $\Delta v = 0.8$ km s$^{-1}$. 

\subsection{Archival data}\label{sec_archival}
 
For the analysis presented here we also used pipeline-processed spectroscopic HARPS archival data\footnote[7]{\url{http://archive.eso.org/wdb/wdb/adp/phase3_main/form}}, acquired between Nov-$2003$ and Dec-$2016$. From all the available observations ($2046$), we considered only the highest S/N spectrum acquired during each night, which led to a total of $60$ additional standard spectra (i.e. not in polarimetric mode). The phase 3 (PH3) data, which correspond to the final data science products of the ESO archive, are fully reduced but no continuum normalization is applied. Therefore, we followed the same two-step normalization procedure as with the spectropolarimetric observations, described in the second part of Sect.~\ref{lv1}.     

\section{Results}\label{sect_4}

\subsection{Radial velocity measurements}\label{sec_RV}

In each individual night we used the extracted LSD Stokes~I profile to measure the RV of the star ($v_{\rm r}$), by considering the resulting centroid from a least-squares Gaussian fit to the data (cf. \citeads{2014MNRAS.444.3517M}, \citeads{2016MNRAS.461.1465H}). As a consistency check, we measure $v_{\rm r}$ in a couple of spectra per epoch with \textit{iSpec} using the default fitting procedure of the Cross-Correlation-Function (CCF), derived from the provided HARPS/SOPHIE G2 customized line mask (\citeads{2014A&A...566A..98B}). Both procedures led to very similar results and indicated a typical RV error of $\sim$$2$ m s$^{-1}$, as expected for HARPS measurements of a F8V-G0V star with $v\sin(i) \simeq 6.5$~km s$^{-1}$ (\citeads{2010exop.book...27L}). 

\subsection{Magnetic activity indicators}\label{sec_activity}

Measurements of two different magnetic activity indicators were performed. The first one corresponds to the HARPS chromospheric S-index ($S_{\rm H}$), defined by the ratio

\begin{equation}\label{eq1}
S_{\rm H} = \alpha\left(\dfrac{H+K}{R+V}\right)\mbox{.}
\end{equation}

\begin{table*}
\caption{Journal of observations (columns $1-5$) and measurements for each night (columns $6-11$).}             
\label{table_2}      
\centering
\begin{threeparttable}
{\scriptsize       
\begin{tabular}{c c c c c c c c r r c}    
\hline\hline
& & & & & & & & &\\[-4pt]

& BJD (+2400000.) & \# Exp.$^{c}$ & \multicolumn{2}{c}{S/N [@551 nm]} & $v_{\rm r}$ & \multicolumn{2}{c}{Activity Indicators} & \multicolumn{1}{c}{$B_{\ell}$} & \multicolumn{1}{c}{$N_{\ell}$} & $\sigma_{\ell}$ \\[3pt]

& [days] & & I & V & [m s$^{-1}$] & $S_{\rm H}$ & $I_{\rm H\alpha}$ & \multicolumn{1}{c}{[G]} & \multicolumn{1}{c}{[G]} & [G] \\[3pt]
\hline
& & & & & & & &\\[-3pt]                    
\parbox[t]{1mm}{\multirow{11}{*}{\rotatebox[origin=c]{90}{\textbf{(Oct. 2015)}}}} & 57300.78580  & 2 & 677 & 657 & 16887.00 $\pm$ 2.53  &   0.2315 $\pm$ 0.0056  &   0.5115 $\pm$ 0.0014  &     2.23  & 0.05 & 0.41\\  
 & 57301.78847  & 1 & 241 & 222 &  16889.58 $\pm$ 4.41  &   0.2281 $\pm$ 0.0190  &   0.5127 $\pm$ 0.0039  &     1.52  &     3.05   &  1.23\\  
 & 57302.78891  & 2 & 559 & 535 &  16907.38 $\pm$ 2.68  &   0.2486 $\pm$ 0.0072  &   0.5124 $\pm$ 0.0017  &     0.21  &     0.01   &  0.49\\  
 & 57303.79205  & 2 & 849 & 810 &  16903.73 $\pm$ 2.33  &   0.2442 $\pm$ 0.0046  &   0.5140 $\pm$ 0.0011  &$-$0.46  &$-$0.01  & 0.32\\   
 & 57304.78161  & 2 & 306 & 230 &  16914.80 $\pm$ 3.82  &   0.2318 $\pm$ 0.0138  &   0.5118 $\pm$ 0.0032  &     0.58  &     0.70   & 1.16\\   
 & 57305.78524  & 2 & 648 & 617 &  16894.02 $\pm$ 2.67  &   0.2460 $\pm$ 0.0062  &   0.5123 $\pm$ 0.0015  &     0.92  &     0.04   & 0.42\\   
 & 57306.78315  & 2 & 814 & 771 &  16898.14 $\pm$ 2.41  &   0.2372 $\pm$ 0.0050  &   0.5124 $\pm$ 0.0012  &$-$2.29  &     0.32   & 0.33\\  
 & 57307.77570  & 2 & 726 & 709 &  16910.59 $\pm$ 2.40  &   0.2355 $\pm$ 0.0054  &   0.5118 $\pm$ 0.0013  &$-$1.64  &     0.11   & 0.37\\  
 & 57308.78898  & 1 & 442 & 425 &  16899.08 $\pm$ 3.21  &   0.2401 $\pm$ 0.0092  &   0.5127 $\pm$ 0.0021  &     1.99  &    $-$0.49   & 0.62\\  
 & 57311.78572  & 2 & 922 & 899 & 16917.38 $\pm$ 2.26  &   0.2449 $\pm$ 0.0043  &   0.5138 $\pm$ 0.0010  &$-$1.30  &    $-$0.18   & 0.29\\   
 & 57312.76899  & 2 & 778 & 752 & 16914.48 $\pm$ 2.35  &   0.2460 $\pm$ 0.0050  &   0.5141 $\pm$ 0.0012  &     0.48  &     0.32   &   0.35\\[3pt]
\parbox[t]{1mm}{\multirow{11}{*}{\rotatebox[origin=c]{90}{\textbf{(Dec. 2015)}}}}  & 57374.55060  & 2 & 1068 & 1025 &  16997.89 $\pm$ 2.12  &   0.2462 $\pm$ 0.0036  &   0.5149 $\pm$ 0.0009  &$-$0.07  & $-$0.71 & 0.26\\  
 & 57375.54055  & 2 &   811 & 769 &  16966.98 $\pm$ 2.39  &   0.2461 $\pm$ 0.0047  &   0.5150 $\pm$ 0.0012  &$-$0.23  &    0.35    &  0.33\\  
 & 57379.55264  & 2 & 1159 & 1108 &  17000.88 $\pm$ 1.98  &   0.2388 $\pm$ 0.0035  &   0.5136 $\pm$ 0.0008  &     1.55  &    0.62     &  0.23\\  
 & 57380.55509  & 2 & 1005 & 964 &  16994.96 $\pm$ 2.15  &   0.2430 $\pm$ 0.0039  &   0.5140 $\pm$ 0.0009  &     3.98  &   $-$0.01     &  0.27\\  
 & 57381.55544  & 2 & 1090 & 1038 &  16994.17 $\pm$ 2.12  &   0.2435 $\pm$ 0.0036  &   0.5141 $\pm$ 0.0009  &     3.00  &     0.14     &  0.25\\ 
 & 57382.55338  & 2 & 1010 & 950 &  16993.68 $\pm$ 2.12  &   0.2439 $\pm$ 0.0039  &   0.5139 $\pm$ 0.0010  &     2.26  &    0.44     &  0.27\\  
 & 57383.55427  & 2 & 1214 & 1136 &  16969.50 $\pm$ 1.96  &   0.2403 $\pm$ 0.0034  &   0.5131 $\pm$ 0.0008  &$-$0.19  &   $-$0.06     &  0.22\\  
 & 57384.55438  & 2 & 1141 & 1071 &  16996.00 $\pm$ 2.04  &   0.2439 $\pm$ 0.0036  &   0.5129 $\pm$ 0.0008  &$-$1.58  &    0.24     &  0.24\\  
 & 57385.55417  & 2 & 1036 & 986 &  17009.91 $\pm$ 2.03  &   0.2464 $\pm$ 0.0038  &   0.5133 $\pm$ 0.0009  &$-$1.37  &    0.33     &  0.26\\  
 & 57386.54757  & 2 & 1100 & 1062 &  17002.64 $\pm$ 2.04  &   0.2452 $\pm$ 0.0036  &   0.5135 $\pm$ 0.0009  &$-$0.37  &    0.13     &  0.24\\  
 & 57387.56851  & 2 & 1083 & 1039 &  17013.08 $\pm$ 2.01  &   0.2504 $\pm$ 0.0040  &   0.5138 $\pm$ 0.0009  &     2.03  &    0.48     &  0.25\\[3pt]  
\parbox[t]{1mm}{\multirow{9}{*}{\rotatebox[origin=c]{90}{\textbf{(Feb. 2016)}}}} & 57436.54183  & 3 & 1156 & 1106 &  16982.76 $\pm$ 2.03  &   0.2522 $\pm$ 0.0042  &   0.5148 $\pm$ 0.0008  &$-$0.58  &  $-$0.09 & 0.25\\  
 & 57437.53818  & 3 & 1306 & 1257 & 16970.25 $\pm$ 1.85  &   0.2540 $\pm$ 0.0039  &   0.5146 $\pm$ 0.0007  &$-$0.21  &   $-$0.10    & 0.21\\  
 & 57438.55301  & 3 & 1112 & 1060 & 16967.33 $\pm$ 2.09  &   0.2472 $\pm$ 0.0043  &   0.5147 $\pm$ 0.0008  &     0.91  &   $-$0.14    &  0.25\\   
 & 57439.54492  & 3 & 1192 & 1157 & 16995.00 $\pm$ 1.94  &   0.2470 $\pm$ 0.0041  &   0.5139 $\pm$ 0.0008  &     1.49  &   $-$0.00    & 0.23\\  
 & 57440.54027  & 3 & 1332 & 1276 & 16950.56 $\pm$ 1.82  &   0.2516 $\pm$ 0.0039  &   0.5132 $\pm$ 0.0007  &     1.39  &     0.01   &  0.21\\  
 & 57441.53922  & 3 & 1040 & 996 & 16931.62 $\pm$ 2.22  &   0.2465 $\pm$ 0.0048  &   0.5126 $\pm$ 0.0009  &     0.08  &   $-$0.31    & 0.27\\   
 & 57442.54183  & 3 &  964 &  929 & 16954.12 $\pm$ 2.16  &   0.2413 $\pm$ 0.0052  &   0.5109 $\pm$ 0.0009  &     0.81  &   $-$0.01    &  0.29\\   
 & 57444.54762  & 2 &  357 &  345 & 16937.70 $\pm$ 3.66  &   0.2604 $\pm$ 0.0126  &   0.5121 $\pm$ 0.0027  &$-$0.51  &   $-$0.06    & 0.76\\  
 & 57445.53731  & 3 &  975 &  958 & 16947.29 $\pm$ 2.15  &   0.2471 $\pm$ 0.0047  &   0.5120 $\pm$ 0.0010  &$-$0.50  &   $-$0.24    & 0.28\\[3pt]  
\parbox[t]{1mm}{\multirow{8}{*}{\rotatebox[origin=c]{90}{\textbf{(Jun. 2016)}}}} & 57562.92231 & 3 & 886 & 857 & 16899.33 $\pm$ 2.24  &   0.2420 $\pm$ 0.0043  &   0.5133 $\pm$ 0.0011  &     0.62  & 0.01 & 0.31\\  
 & 57567.92469  & 2 & 880 & 857 & 16887.04 $\pm$ 2.20  &   0.2396 $\pm$ 0.0044  &   0.5129 $\pm$ 0.0011  &     2.07  &    0.36    &  0.31\\   
 & 57568.92232  & 2 & 749 & 712 & 16878.23 $\pm$ 2.53  &   0.2382 $\pm$ 0.0053  &   0.5124 $\pm$ 0.0013  &     1.21  &    0.51   &  0.36\\  
 & 57569.92692  & 2 & 830 & 806 & 16893.89 $\pm$ 2.30  &   0.2441 $\pm$ 0.0045  &   0.5118 $\pm$ 0.0012  &$-$1.10  &   $-$0.10   &  0.32\\  
 & 57570.91827  & 2 & 852 & 825 & 16884.99 $\pm$ 2.32  &   0.2427 $\pm$ 0.0045  &   0.5126 $\pm$ 0.0011  &     0.06  &    0.05   &  0.32\\  
 & 57575.92260  & 2 & 690 & 671 & 16883.47 $\pm$ 2.54  &   0.2461 $\pm$ 0.0056  &   0.5138 $\pm$ 0.0014  &     0.67  &   0.35    &  0.39\\  
 & 57576.92258  & 2 & 738 & 706 & 16881.69 $\pm$ 2.55  &   0.2444 $\pm$ 0.0053  &   0.5130 $\pm$ 0.0013  &     0.05  &  $-$0.35    &  0.37\\   
 & 57589.85636  & 3 & 889 & 861 & 16895.13 $\pm$ 2.29  &   0.2388 $\pm$ 0.0043  &   0.5117 $\pm$ 0.0011   &    1.06  &   $-$0.11    &  0.30\\[3pt]  
\parbox[t]{1mm}{\multirow{11}{*}{\rotatebox[origin=c]{90}{\textbf{(Aug. 2016)}}}} & 57619.81856  & 2 & 618 & 593 & 16905.42 $\pm$ 2.80  &   0.2472 $\pm$ 0.0063  &   0.5139 $\pm$ 0.0015  &$-$0.97  &  $-$0.02 & 0.45\\  
 & 57621.82170  & 2 &  719 & 693 & 16899.17 $\pm$ 2.49  &   0.2396 $\pm$ 0.0053  &   0.5137 $\pm$ 0.0013  &     0.26  &    0.31    &  0.37\\  
 & 57622.83422  & 2 &  781 & 751 & 16905.58 $\pm$ 2.48  &   0.2397 $\pm$ 0.0049  &   0.5135 $\pm$ 0.0012  &     0.54  &    0.51    &  0.34\\  
 & 57623.81619  & 2 &  955 & 929 & 16914.81 $\pm$ 2.12  &   0.2402 $\pm$ 0.0041  &   0.5128 $\pm$ 0.0010  &     0.70  &    0.09    &  0.28\\  
 & 57624.82099  & 2 &  784 & 753 & 16908.49 $\pm$ 2.43  &   0.2427 $\pm$ 0.0050  &   0.5139 $\pm$ 0.0012  &     0.97  &   $-$0.80    &  0.35\\  
 & 57625.82139  & 2 &  977 & 941 & 16937.99 $\pm$ 2.20  &   0.2457 $\pm$ 0.0040  &   0.5141 $\pm$ 0.0010  &     2.16  &    0.38    &   0.27\\  
 & 57626.86328  & 2 &  566 & 538 & 16928.33 $\pm$ 2.81  &   0.2489 $\pm$ 0.0064  &   0.5151 $\pm$ 0.0017  &     0.79  &    0.00    &  0.47\\  
 & 57629.81840  & 2 &  862 & 856 & 16905.51 $\pm$ 2.23  &   0.2428 $\pm$ 0.0044  &   0.5130 $\pm$ 0.0011  &$-$0.16  &   $-$0.17    &  0.31\\  
 & 57630.82050  & 2 &  974 & 929 & 16919.88 $\pm$ 2.18  &   0.2419 $\pm$ 0.0039  &   0.5135 $\pm$ 0.0010  &$-$0.30  &   $-$0.05    &  0.27\\  
 & 57631.81518  & 2 & 1118 & 1069 & 16921.00 $\pm$ 1.98  &   0.2441 $\pm$ 0.0035  &   0.5133 $\pm$ 0.0009  &     0.28  &   $-$0.03    &  0.24\\   
 & 57632.81974  & 2 &  904 & 868 & 16914.84 $\pm$ 2.29  &   0.2435 $\pm$ 0.0043  &   0.5141 $\pm$ 0.0011  &     0.48  &    $-$0.27    &  0.30\\[3pt] 
\parbox[t]{1mm}{\multirow{14}{*}{\rotatebox[origin=c]{90}{\textbf{(Sep. 2016)}}}} & 57645.79605  & 4 & 948 & 916 & 16981.41 $\pm$ 2.25  &   0.2467 $\pm$ 0.0040  &   0.5144 $\pm$ 0.0010  &     0.79  & 0.02 & 0.29\\  
 & 57646.77036  & 2 &   739 &   678 & 16949.99 $\pm$ 2.40  &   0.2465 $\pm$ 0.0050  &   0.5134 $\pm$ 0.0013  &     0.06  &    0.38    &  0.38\\   
 & 57647.78723  & 2 &  1107 & 1081 & 16947.88 $\pm$ 1.96  &   0.2459 $\pm$ 0.0035  &   0.5131 $\pm$ 0.0009  &$-$0.08  &    0.36    &  0.24\\   
 & 57648.81725  & 2 &   998 &   961 & 16945.52 $\pm$ 2.03  &   0.2522 $\pm$ 0.0038  &   0.5134 $\pm$ 0.0010  &     3.00  &    0.12    &  0.27\\  
 & 57649.84913  & 2 &   735 &   707 & 16935.93 $\pm$ 2.47  &   0.2509 $\pm$ 0.0049  &   0.5140 $\pm$ 0.0013  &     3.24  &    0.14    &  0.36\\  
 & 57650.85027  & 2 &  1092 & 1032 & 16921.65 $\pm$ 2.01  &   0.2458 $\pm$ 0.0037  &   0.5142 $\pm$ 0.0009  &     1.17  &    0.07    &  0.25\\  
 & 57651.84535  & 2 &   953 &   924 & 16952.32 $\pm$ 2.16  &   0.2442 $\pm$ 0.0041  &   0.5136 $\pm$ 0.0010  &$-$0.41  &    0.08    &  0.28\\   
 & 57652.85980  & 2 &  1037 &  996 & 16948.99 $\pm$ 1.93  &   0.2465 $\pm$ 0.0039  &   0.5137 $\pm$ 0.0009  &     0.50  &    0.38    &  0.26\\  
 & 57654.84500  & 3 &   790 &   771 & 16958.12 $\pm$ 2.34  &   0.2421 $\pm$ 0.0049  &   0.5132 $\pm$ 0.0012  &$-$1.38  &   $-$0.12    &  0.34\\  
 & 57655.86385  & 3 &  1133 & 1094 & 16974.67 $\pm$ 1.96  &   0.2453 $\pm$ 0.0039  &   0.5132 $\pm$ 0.0008  &     0.98  &   $-$0.16    &  0.24\\  
 & 57656.80459  & 2 &  1073 & 1051 & 16946.08 $\pm$ 2.00  &   0.2475 $\pm$ 0.0037  &   0.5144 $\pm$ 0.0009  &     1.88  &    0.00    &  0.25\\  
 & 57657.84229  & 2 &   960 &   928 & 16939.42 $\pm$ 2.10  &   0.2531 $\pm$ 0.0047  &   0.5146 $\pm$ 0.0010  &     2.53  &    0.09    &  0.29\\   
 & 57658.82478  & 2 &  1117 &  1071 &16947.05 $\pm$ 2.02  &   0.2483 $\pm$ 0.0037  &   0.5142 $\pm$ 0.0009  &$-$0.56  &    0.51    &  0.24\\  
 & 57659.82063  & 2 &   841 &   796 & 16973.45 $\pm$ 2.25  &   0.2470 $\pm$ 0.0045  &   0.5139 $\pm$ 0.0012  &$-$2.00  &   $-$0.10    &  0.32\\[3pt]  
\parbox[t]{1mm}{\multirow{11}{*}{\rotatebox[origin=c]{90}{\textbf{(Oct. 2016)}}}}  & 57676.78694  & 2 & 595 & 576 & 16980.22 $\pm$ 2.76  &   0.2520 $\pm$ 0.0072  &   0.5129 $\pm$ 0.0016  &  2.09  & 1.28 & 0.48\\  
 & 57679.83434  & 1 &  330  & 310  & 16945.60 $\pm$ 3.87  &   0.2418 $\pm$ 0.0129  &   0.5138 $\pm$ 0.0029  &$-$0.14  & 2.29&    0.84\\  
 & 57680.81105  & 2 &  506  & 500 & 16962.62 $\pm$ 2.97  &   0.2355 $\pm$ 0.0079  &   0.5135 $\pm$ 0.0019  &     1.70 &  1.00 &   0.53 \\  
 & 57681.79977  & 2 &  825  & 787  & 16974.86 $\pm$ 2.40  &   0.2462 $\pm$ 0.0051  &   0.5136 $\pm$ 0.0011  &$-$1.45  & 0.10 &   0.33 \\  
 & 57682.81243  & 2 & 1015 & 979   & 16993.23 $\pm$ 2.07  &   0.2448 $\pm$ 0.0043  &   0.5143 $\pm$ 0.0009  &    0.60  &$-$0.22 &   0.27 \\  
 & 57683.78929  & 1 &  398  & 377  & 16962.76 $\pm$ 3.40  &   0.2456 $\pm$ 0.0102  &   0.5135 $\pm$ 0.0024  &     0.60  & 0.20 & 0.69 \\  
 & 57684.75537  & 2 &  618  & 588  & 16983.77 $\pm$ 2.76  &   0.2449 $\pm$ 0.0063  &   0.5141 $\pm$ 0.0015  &     1.83  &$-$0.04 &   0.44  \\  
 & 57686.73238  & 2 &  582  & 564  & 16986.12 $\pm$ 2.79  &   0.2463 $\pm$ 0.0067  &   0.5146 $\pm$ 0.0016  &     2.39  & 0.52 & 0.46\\ 
 & 57687.78773  & 2 &  754  & 769  & 16972.72 $\pm$ 2.50  &   0.2440 $\pm$ 0.0053  &   0.5143 $\pm$ 0.0012  &     2.24  & 0.02 &   0.35 \\   
 & 57688.78882  & 2 &  921  & 870  & 16970.63 $\pm$ 2.28  &   0.2418 $\pm$ 0.0047  &   0.5138 $\pm$ 0.0010  &$-$0.37  & 0.08 &    0.30 \\  
 & 57689.79347  & 2 &  848  & 811  & 16983.95 $\pm$ 2.38  &   0.2430 $\pm$ 0.0053  &   0.5141 $\pm$ 0.0011  &     0.00 & $-$0.09 &   0.33  \\[3pt]  
\hline                 
\end{tabular}}
\begin{tablenotes}{\vspace{-2pt}}
\item {\scriptsize $^{(c)}$ Number of Stokes V exposures used in the analysis.}\\
\smallskip
\item {\hfill \textit{Continued on next page.}}
\end{tablenotes}
\end{threeparttable}
\end{table*}

\begin{table*}
\begin{threeparttable}
{\hfill \textit{Continued from previous page.}}
\smallskip
\smallskip
{\scriptsize       
\begin{tabular}{c c c c c c c c r r c}    
\hline\hline
& & & & & & & & &\\[-4pt]

& BJD (+2400000.) & \# Exp. & \multicolumn{2}{c}{S/N [@551 nm]} & $v_{\rm r}$ & \multicolumn{2}{c}{Activity Indicators} & \multicolumn{1}{c}{$B_{\ell}$} & \multicolumn{1}{c}{$N_{\ell}$} & $\sigma_{\ell}$ \\[3pt]

& [days] & & I & V & [m s$^{-1}$] & $S_{\rm H}$ & $I_{\rm H\alpha}$ & \multicolumn{1}{c}{[G]} & \multicolumn{1}{c}{[G]} & [G] \\[3pt]
\hline
& & & & & & & &\\[-3pt]                    
\parbox[t]{1mm}{\multirow{12}{*}{\rotatebox[origin=c]{90}{\textbf{(Dec. 2016)}}}} & 57737.64577  & 2 & 901 & 864 & 16968.45 $\pm$ 2.33  &   0.2504 $\pm$ 0.0049  &   0.5128 $\pm$ 0.0010  &$-$2.39  & $-$0.06 & 0.48\\   
 & 57740.65620  & 2 & 1009  & 970 & 16968.99 $\pm$ 2.14  &   0.2528 $\pm$ 0.0046  &   0.5152 $\pm$ 0.0009  & $-$0.22 &   $-$0.41    &  0.27\\   
 & 57741.67881  & 2 & 1002  & 966 & 16961.69 $\pm$ 2.17  &   0.2500 $\pm$ 0.0048  &   0.5150 $\pm$ 0.0009  &       0.62 &  $-$0.32    &  0.28\\  
 & 57742.66953  & 2 & 1018  & 968 & 16967.62 $\pm$ 2.11  &   0.2547 $\pm$ 0.0048  &   0.5150 $\pm$ 0.0009  &       1.83 &  $-$0.20     & 0.27 \\  
 & 57743.67149  & 2 &  984  &  938 &16949.96 $\pm$ 2.20  &   0.2535 $\pm$ 0.0046  &   0.5152 $\pm$ 0.0010  &        1.95 &   0.27         & 0.28 \\  
 & 57744.64682  & 2 &  978  &  938 &16956.08 $\pm$ 2.17  &   0.2510 $\pm$ 0.0045  &   0.5147 $\pm$ 0.0010  &  $-$0.62  &   0.13         & 0.28 \\   
 & 57745.65298  & 2 &  441  &  373 &16966.66 $\pm$ 3.28  &   0.2409 $\pm$ 0.0106  &   0.5140 $\pm$ 0.0022  &  $-$3.25 &    0.22         & 0.71 \\  
 & 57746.66875  & 2 & 1074 & 1039 & 16965.44 $\pm$ 2.00  &   0.2547 $\pm$ 0.0047  &   0.5138 $\pm$ 0.0009 & $-$2.79  &    0.29         & 0.26 \\  
 & 57747.66166  & 2 &  899  &  876 &16953.07 $\pm$ 2.33  &   0.2579 $\pm$ 0.0053  &   0.5151 $\pm$ 0.0010  &  $-$1.81 &   $-$0.04     & 0.31 \\  
 & 57748.66862  & 2 &  561  &  539 &16956.91 $\pm$ 2.89  &   0.2510 $\pm$ 0.0076  &   0.5144 $\pm$ 0.0017  &        0.17 &   0.26         & 0.49 \\  
 & 57749.67434  & 2 &  680  &  658 &16941.97 $\pm$ 2.56  &   0.1827 $\pm$ 0.0347  &   0.5147 $\pm$ 0.0013  &  $-$0.81 &    0.04         & 0.49 \\   
 & 57750.66759  & 2 &  823  &  784 &16942.47 $\pm$ 2.36  &   0.2527 $\pm$ 0.0054  &   0.5142 $\pm$ 0.0011  &        1.77 &  $-$0.47     & 0.34 \\[3pt]  
\parbox[t]{1mm}{\multirow{14}{*}{\rotatebox[origin=c]{90}{\textbf{(Feb. 2017)}}}} & 57781.56521  & 2 & 711 & 680 & 16897.54 $\pm$ 2.67  &   0.2479 $\pm$ 0.0064  &   0.5130 $\pm$ 0.0013  &  2.05  & 0.57 & 0.40\\  
 & 57782.57006  & 2 &  705  & 662 & 16904.26 $\pm$ 2.62  &   0.2429 $\pm$ 0.0064  &   0.5134 $\pm$ 0.0013  & 0.89      &  0.04        & 0.39\\  
 & 57783.56522  & 2 &  932  & 880 & 16901.47 $\pm$ 2.22  &   0.2499 $\pm$ 0.0053  &   0.5128 $\pm$ 0.0010  &$-$3.33  & $-$0.20   &  0.31\\  
 & 57784.56808  & 2 &  778  & 751 & 16910.32 $\pm$ 2.48  &   0.2446 $\pm$ 0.0058  &   0.5135 $\pm$ 0.0012  &$-$3.88  &  0.11        & 0.36\\  
 & 57786.56736  & 2 &  996  & 941 & 16889.40 $\pm$ 2.20  &   0.2402 $\pm$ 0.0052  &   0.5129 $\pm$ 0.0009  &$-$0.93  & $-$0.25   &  0.28\\  
 & 57789.54820  & 2 &  835  & 789 & 16873.32 $\pm$ 2.50  &   0.2470 $\pm$ 0.0056  &   0.5144 $\pm$ 0.0011  &  1.54     &  0.09       &   0.33\\  
 & 57790.54995  & 2 &  931  & 898 & 16893.70 $\pm$ 2.29  &   0.2445 $\pm$ 0.0050  &   0.5130 $\pm$ 0.0010  & 1.85      & $-$0.17  &   0.30\\  
 & 57791.55658  & 2 &  848  & 812 & 16895.03 $\pm$ 2.35  &   0.2438 $\pm$ 0.0054  &   0.5139 $\pm$ 0.0011  &$-$2.06  &  0.37       &  0.33\\  
 & 57793.56295  & 2 &  921  & 883 & 16914.98 $\pm$ 2.18  &   0.2441 $\pm$ 0.0054  &   0.5141 $\pm$ 0.0010  &$-$0.24  & $-$0.13  &   0.30\\  
 & 57794.56707  & 2 &  852  & 810 & 16896.89 $\pm$ 2.35  &   0.2444 $\pm$ 0.0056  &   0.5145 $\pm$ 0.0011  &$-$0.86  &  0.05      &   0.33\\  
 & 57795.56821  & 2 &  706  & 673 & 16920.12 $\pm$ 2.71  &   0.2462 $\pm$ 0.0065  &   0.5145 $\pm$ 0.0013  &$-$1.35  &  0.22      &   0.39\\  
 & 57796.56822  & 2 &  993  & 925 & 16904.89 $\pm$ 2.07  &   0.2450 $\pm$ 0.0052  &   0.5140 $\pm$ 0.0009  &$-$1.15  & $-$0.05   &  0.28\\  
 & 57797.56245  & 2 & 1071 &1023 &  16878.10 $\pm$ 2.18  &   0.2430 $\pm$ 0.0049  &   0.5130 $\pm$ 0.0009  & 1.98     &  0.23       &  0.26\\
 & 57798.57220  & 2 &  921  & 879 & 16896.94 $\pm$ 2.25  &   0.2423 $\pm$ 0.0055  &   0.5127 $\pm$ 0.0010  & 0.21      &  0.07       &  0.31\\[1pt]

\hline                  
\end{tabular}}
\end{threeparttable}
\end{table*}

\noindent In this expression, $H$ and $K$ represent the fluxes measured in $0.105$ nm bandpasses, centered at the line cores of the of the Ca II H \& K lines (located at $396.8492$ nm and $393.3682$~nm, respectively). Similarly, $R$ and $V$ correspond to the integrated fluxes in two nearby continuum regions spanning $2$~nm, and centered at $390.107$ nm and $400.107$ nm, respectively. To convert $S_{\rm H}$ to the classical Mt. Wilson scale we used the conversion factor $\alpha = 15.39 \pm 0.65$, previously determined by \citetads{2015A&A...582A..38A}.

The second activity indicator considered is the H$\alpha$-index ($I_{\rm H\alpha}$), obtained from the line to continuum flux ratio

\begin{equation}\label{eq2}
I _{\rm H\alpha} = \dfrac{F_{\rm H\alpha}}{C_{\rm B}+C_{\rm R}}\mbox{.}
\end{equation}

\noindent Similar to the S-index, $F_{\rm H\alpha}$ denotes the flux integrated in a $0.36$ nm bandpass around the H$\alpha$ line core (at $656.2801$~nm), while $C_{\rm B}$ and $C_{\rm R}$ are the fluxes in two $0.22$ nm pseudo-continuum regions, with central wavelengths located towards the blue ($655.885$ nm) and red ($656.730$ nm) of the H$\alpha$ line (cf. \citeads{2002AJ....123.3356G}, \citeads{2014MNRAS.444.3517M}). Similar definitions for $I_{\rm H\alpha}$ have been employed in previous activity-RV studies of M-dwarf stars, using narrower bandpasses for the line core and continuum flux integration (e.g. \citeads{2003A&A...403.1077K}, \citeads{2007A&A...474..293B}).

\subsection{Longitudinal magnetic field}\label{sec_bl}

Unlike the activity indicators discussed in the previous section, measurements of the surface-averaged longitudinal magnetic field ($B_{\ell}$) are performed using the results of the LSD analysis (Sect.~\ref{lv2}). For a pair of LSD Stokes $I(v)$ and $V(v)$ profiles, $B_{\ell}$ can be estimated (in Gauss) from the following expression (\citeads{1997MNRAS.291..658D}, \citeads{2009ARA&A..47..333D}):

\begin{equation}\label{eq3}
B _{\ell} = -714\dfrac{\int v V(v){\rm d}v}{\lambda_0\bar{g}\int\left[I_{\rm c} - I(v)\right]{\rm d}v}\mbox{.}
\end{equation}

\noindent Here, $v$ is the velocity coordinate (in units of km s$^{-1}$) and $I_{\rm c}$ represents the continuum intensity (in normalized units). The remaining parameters correspond to the central wavelength ($\lambda_0 \simeq 0.509$ $\muup$m), and mean Land\'e factor ($\bar{g}~\simeq~1.198$), of the extracted LSD Stokes I profiles. The uncertainties derived from the spectra through the LSD procedure are propagated to estimate the errors in $B _{\ell}$. However, as discussed by \citetads{2014MNRAS.444.3517M}, additional errors arise from the integration limits used in eq. \ref{eq3}. These need to be wide enough to cover the entire profiles, without over extending into the noise-dominated continuum. Therefore, we adopted a velocity range between $5.4$ km s$^{-1}$ and $30.2$ km s$^{-1}$ which maximizes the $B _{\ell}/\sigma_{\ell}$ ratio, with $\sigma_{\ell}$ as the final error of the measurement.     

Finally, similar measurements (denoted as $N_{\ell}$) are performed by replacing $V(v)$ in eq. \ref{eq3} for the LSD $N(v)$ profiles. This quantity provides a first-order indication of whether $B_{\ell}$ may have contributions from spurious polarization (i.e.~when $N_{\ell} > 3\sigma_{\ell}$). Values of $N_{\ell} \simeq 0$ are expected (see Sect. \ref{lv1}), and typically correspond to robust $B_{\ell}$ measurements. However, as these are integrated quantities it is necessary to visually inspect the LSD Stokes $V(v)$ and $N(v)$ profiles, since it is possible to obtain null measurements of either $B_{\ell}$ or $N_{\ell}$ from profiles where a clear signature is visible. 

The resulting values of $B_{\ell}$,  $N_{\ell}$, and $\sigma_{\ell}$, as well as the measurements described in Sects. \ref{sec_RV} and \ref{sec_activity}, are listed in Table \ref {table_2} (columns $6 - 11$).

\section{Analysis and Discussion}\label{sect_5}

\subsection{Evolution of the activity: multiple cycles?}\label{sec_cycles}

\noindent The first part of the analysis considers the long-term evolution in the activity levels of $\iota$ Hor. We will focus our discussion on the Ca H\,\&\,K indicator, as the H$\alpha$ index follows a very similar trend (albeit with a much lesser degree of variability), and there are no previous reports in the literature of the $I_{\rm H\alpha}$ indicator for this particular star. 

\begin{figure*}
\includegraphics[width=\linewidth]{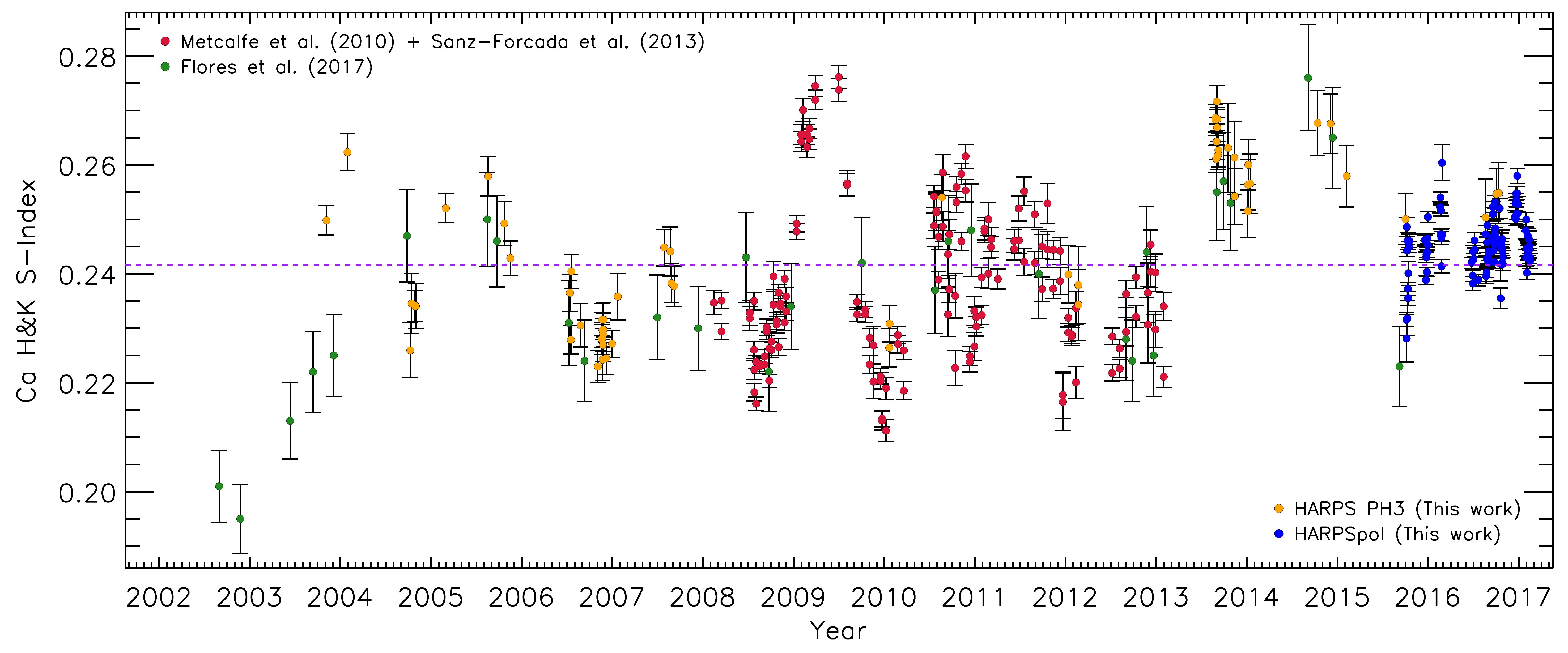}
\caption{Long-term evolution of the chromospheric Ca H\,\&\,K S-Index of $\iota$ Hor.}\label{fig_1}
\end{figure*}

\noindent Figure \ref{fig_1} shows our S-index measurements, extracted from the HARPS PH3 (Sect.~\ref{sec_archival}) and HARPSpol datasets, alongside values previously reported by \citetads{2010ApJ...723L.213M}, \citetads{2013A&A...553L...6S}, and \citetads{2017MNRAS.464.4299F}. The former two datasets served to identify the $1.6$-yr activity cycle, while the latter\footnote[8]{The analysis of \citetads{2017MNRAS.464.4299F} also considered monthly-averages derived from HARPS archival data. Those values are not included in Fig. \ref{fig_1}.} was used to obtain a secondary period of $\sim$$4.5$-yr. By considering the combined dataset, it is clear that the activity cycles of $\iota$ Hor are relatively irregular and have episodes where no obvious periodicity is visible in the data.

As shown in the figure, our HARPS PH3 S-index values closely follow the previously reported contemporaneous measurements, which were obtained using two different instruments with very disimilar spectral resolutions: RC~Spec ($R \sim 2500$) in the case of \citetads{2010ApJ...723L.213M} and \citetads{2013A&A...553L...6S}, and REOSC ($R \sim 26400$) in the case of \citetads{2017MNRAS.464.4299F}. This indicates that our HARPS calibration to the Mt. Wilson scale is robust, and gives confidence to the $S_{\rm H}$ values determined at times where no additional observations are available (i.e.~the~HARPSpol epochs). 
 

\begin{figure} 
\includegraphics[trim=0.55cm 0.8cm 0.83cm 0.69cm, clip=true, width=\linewidth]{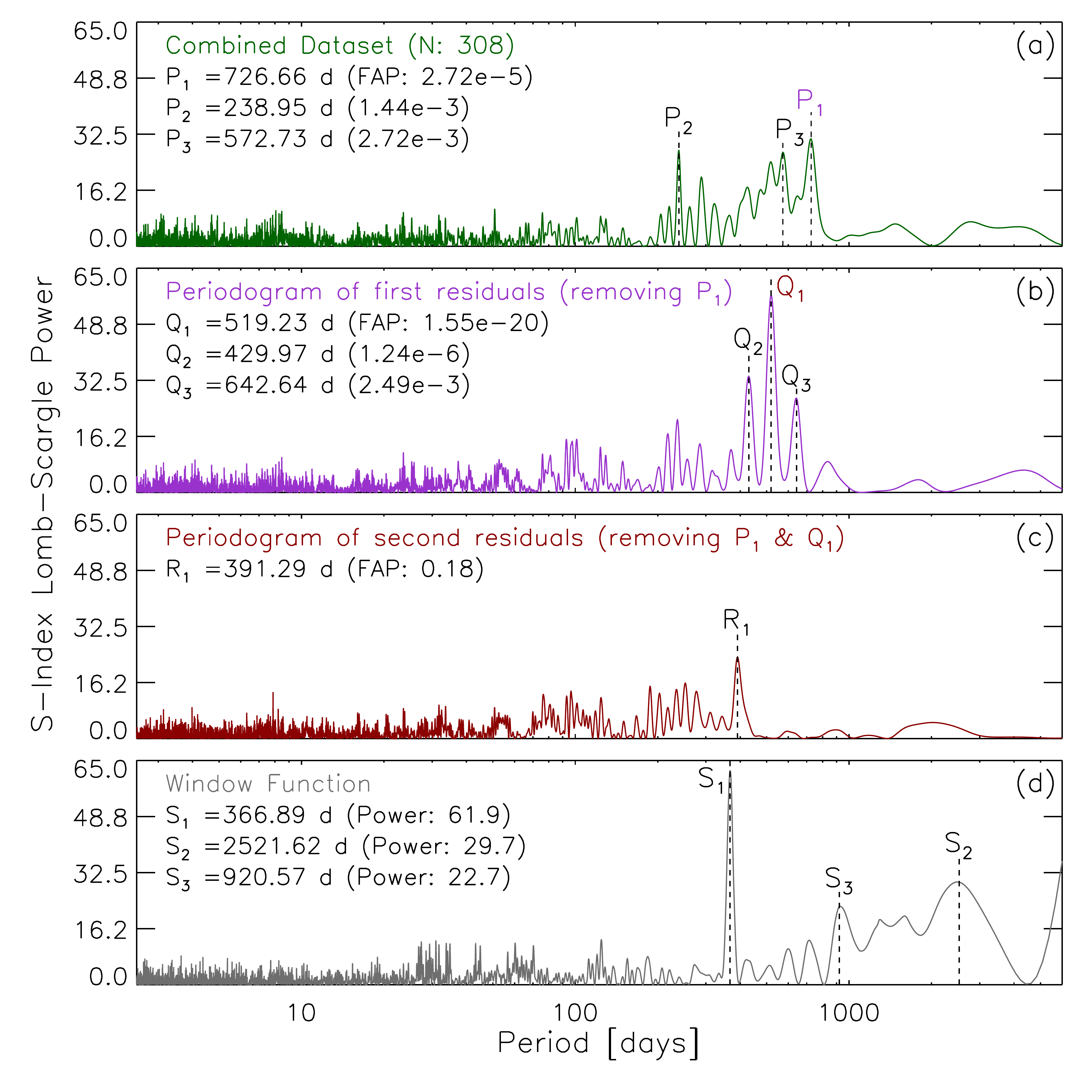}
\caption{Results from periodicity analysis on the long-term S-index variations of $\iota$ Hor (Fig. \ref{fig_1}). From \textit{top} to \textit{bottom}, Lomb-Scargle (LS) periodograms are presented for the original dataset, the first and second residuals, and the window function.}\label{fig_2}
\end{figure}

Even though our $101$ new data points from HARPSpol significantly increase the number of available S-index measurements, their $\sim$$1.4$ yr time-span is not sufficient to consider them alone for cycle determination purposes. For this reason we used the combined dataset, consisting of $334$ individual measurements spanning $\sim$$14$ yr, to check for periodic signals via the classical Lomb-Scargle (LS) periodogram analysis (\citeads{1976Ap&SS..39..447L}, \citeads{1982ApJ...263..835S}). 

Panel \textit{a} in Fig. \ref{fig_2} contains the resulting LS periodogram for the combined time-series shown in Fig. \ref{fig_1}, following the power definition given by \citetads{1986ApJ...302..757H}. The periods ($P_{1}$, $P_{2}$, $P_{3}$) corresponding to the three most prominent peaks are presented. The associated False Alarm Probabilities (FAP) have been estimated using a standard bootstrapping randomization algorithm (\citeads{1990Natur.348..407B}), employing $10^5$ synthetic periodograms (preserving the time-spacing of the data points). While the $1.6$ yr ($\sim$$584$ d) periodicity could possibly be associated with the third identified peak ($P_{3} = 572.73$ d), the first peak located at $P_{1} = 726.66$ d has a much lower FAP (by two orders of magnitude). Still, the amplitudes of both of these peaks are comparable, indicating a similar contribution to the periodicity in the time-series. 

\begin{figure*}
\includegraphics[width=\linewidth]{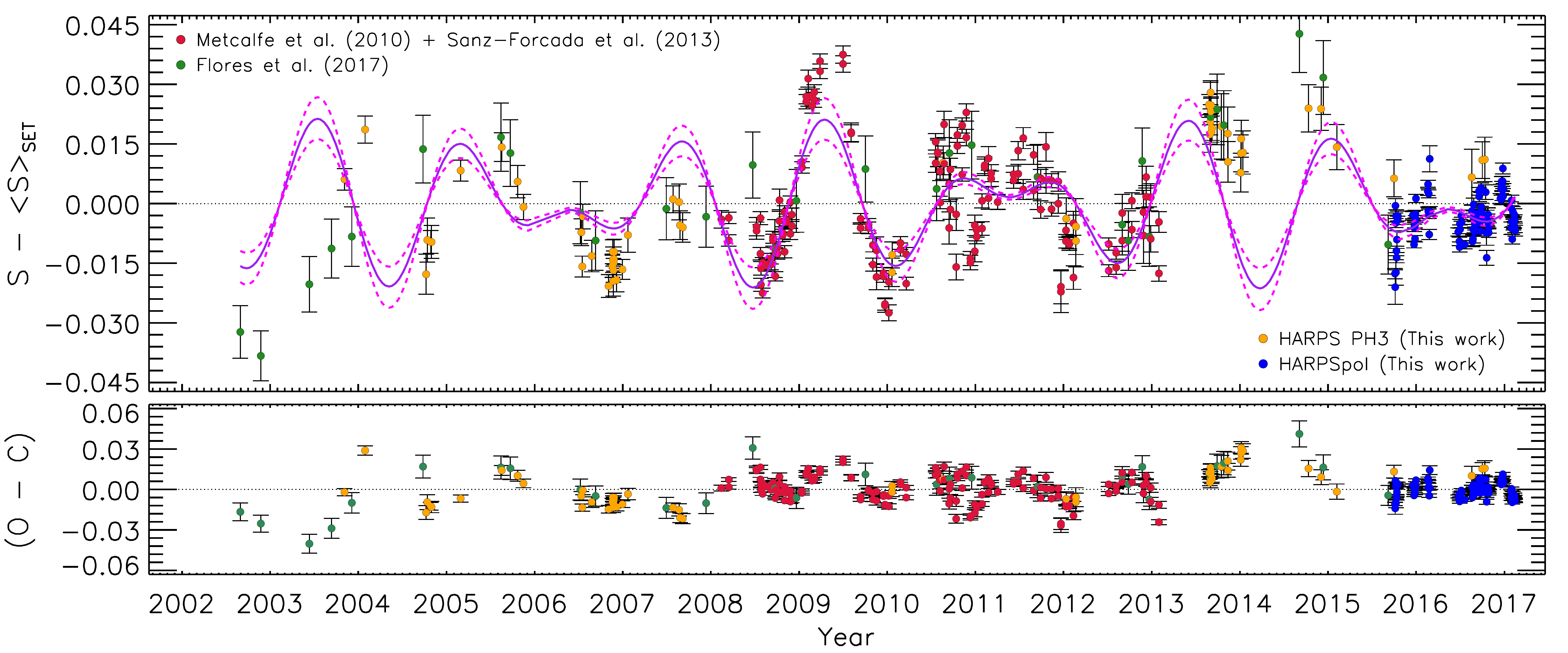}
\caption{Results from the multi-parametric fit to the activity cycle of $\iota$ Hor. The \textit{top panel} contains the variation of the S-index relative to the mean value in each instrument ($S - \left<S\right>_{\rm SET}$). The solid line corresponds to the best 2-period model ($S_{\rm CYC}$), with $P_{1} \simeq 1.97$~yr and $P_{2} \simeq 1.41$~yr ($\chi^2_{\rm r} = 16.73$). The dashed lines show the 3-$\sigma$ cycle amplitude limits, keeping the remaining parameters of the periodic signals fixed. The residuals are presented in the \textit{bottom panel}.}\label{fig_3}
\end{figure*}

To investigate the possibility of multiple periodicities, we consider a multi-parametric fit to the long-term evolution of the chromospheric activity of $\iota$ Hor. In this approach we characterize each signal by a sinusoidal model, composed of three free parameters representing the period ($P$), semi-amplitude ($A$), and phase ($\Phi$) of the modulation. While the entire time-series is used in the analysis, an independent offset is calculated for data sets acquired with different instruments. Therefore, four\footnote[9]{For this treatment we consider the HARPS PH3 and HARPSpol data sets independently.} additional free parameters are included in our final model, corresponding to the mean activity levels $\left<S\right>_{\rm SET}$ (compatible with the best-fit model), associated with a particular data set or instrument.

We follow a sequential procedure to identify and extract the periodicities in a hierarchical manner. In this way, we consider the highest peak of the initial LS periodogram ($P_1$~in Fig.~\ref{fig_2}, panel \textit{a}), and obtain the best-fit multi-parametric model under a reduced $\chi^2_{\rm r}$ minimization scheme. This initial model is then subtracted from the original time-series, from where a LS periodogram of the residuals is computed (Fig. \ref{fig_2}, panel \textit{b}). The process is repeated for the highest peak in this new LS periodogram ($Q_{1}$ in Fig.~\ref{fig_2}, panel \textit{b}), which is then included in the multi-parametric model and fitted simultaneously with the previously defined periodicity (therefore, the fit parameters identified in the first step change slightly). Once again the best-fit model (which now includes 2 periodic signals), is subtracted from the data and a second set of residuals is obtained. Similarly, a LS periodogram of the second residuals is generated (Fig.~\ref{fig_2}, panel \textit{c}), to check if there is any significant signal remaining in the data. As a consistency check, the best-fit periods identified at every step in this process are compared against the location of the resulting peaks of the window function (Fig. \ref{fig_2}, panel \textit{d}).

\begin{table}
\caption{Final 2-period model ($S_{\rm CYC}$) for the activity evolution of $\iota$ Hor.}
\label{tab_3}      
\centering                          
\begin{threeparttable}
\begin{tabular}{l c}        
\hline\hline                 
Parameter & Value \\
\hline
& \\[-8pt]
\textit{1st Component} & \\
Period ($P_1$) & 719 $\pm$ 7 d (1.968 $\pm$ 0.019 yr)\\ 
Semi-amplitude ($A_1$) & 0.0118 $\pm$ 0.0010 \\
Phase ($\Phi_1$) & 6.27 $\pm$ 0.16 rad \\[3pt]
\textit{2nd Component} & \\ 
Period ($P_2$) & 516 $\pm$ 4 d (1.412 $\pm$ 0.010 yr)\\ 
Semi-amplitude ($A_2$) & 0.0102 $\pm$ 0.0008 \\
Phase ($\Phi_2$) & 4.51 $\pm$ 0.04 rad \\[3pt]
\textit{Offsets} $\left<S\right>_{\rm SET}$ & \\ 
REOSC (30)$^{a}$ & 0.233 $\pm$ 0.003 \\ 
RC Spec (143) & 0.2387 $\pm$ 0.0009 \\
HARPS PH3 (60) & 0.244 $\pm$ 0.002 \\  
HARPSpol (101) & 0.2491 $\pm$ 0.0008 \\[3pt] 
\textit{Statistics} & \\ 
Reduced $\chi^2_{\rm r}$ & 16.73 \\ 
Data points  & 334 \\
$\left<S\right>_{\rm ALL}$ & 0.24159 $\pm$ 0.00072\\[1pt]
\hline
\end{tabular}
\begin{tablenotes}
{\small $^{(a)}$ Number of data points for a given instrument.}
\end{tablenotes}
\end{threeparttable}
\end{table}

As presented in panel \textit{c} of Fig. \ref{fig_2}, the LS periodogram of second residuals yielded a peak at $391.29$ d with a FAP~=~$0.18$, which is not considered as significant and indicates the end of our iterative process. In this way, only two significant peaks were identified and for this reason, the multi-parametric model considers two periodic signals (i.e.,~$10$ free parameters). The final best-fit solution ($S_{\rm CYC}$) is compared with the observations in Fig.~\ref{fig_3}, and a summary of the model is listed in Table \ref{tab_3}. The errors have been estimated following a bootstrap resampling scheme (Press et al.~\citeyearads{1992nrca.book.....P}, \citeyearads{2002nrca.book.....P}), generating $10^4$ synthetic time-series (randomly drawn from the original data set), and obtaining a best-fit model for each one of them (the final model to the real data is used as an initial guess to accelerate the process). A 68 per cent confidence interval of the distribution resulting for each parameter of the synthetic fits, is taken as the corresponding error in the final model.       

As can be seen in Fig. \ref{fig_3}, our model reproduces to a very good extent the long-term evolution of the chromospheric activity in the star. The overall behaviour clearly resembles a \textit{beating} pattern, driven by the relatively small differences between the periods and amplitudes of the dominant signals (see Table \ref{tab_3}). Note that none of the chromospheric cycle periods reported in the literature appear in our final solution. As explained below, these discrepancies were expected and can be explained by considering some differences between these previous studies and our analysis. 

As mentioned before, the $1.6$ yr cycle was determined using observations acquired between $2008$ and $2013$ (\citeads{2010ApJ...723L.213M}, \citeads{2013A&A...553L...6S}), where good agreement was obtained towards the beginning and the end of this time span. However, this periodicity is unable to reproduce the entire $\sim$$5$ yr base line, and is particularly inconsistent with the observed activity evolution between late-$2010$ and mid-$2011$. This is no longer the case in our approach, provided that all the observations are considered together and this particular data set puts very strong constraints during the fitting process since it provides the best cycle coverage of the entire time series. In addition, the apparent lack of coherence in the activity cycle observed between late-$2010$ and mid-$2011$ can now be understood in the framework of our model as simply the result of destructive interference between two, out of phase, periodic signals. Furthermore, the typical $\sim$$5$ year\footnote[10]{Formally, this comes from the inverse of the beating frequency $\nu_{\rm beat} = \left|\nu_1 - \nu_2\right| = \left| \frac{1}{P_{1}} - \frac{1}{P_{2}}\right| \simeq \left| \frac{1}{1.97} - \frac{1}{1.41}\right| \simeq~0.2$ yr$^{-1}$.} time-scale of this beating pattern can also explain the secondary cycle proposed by \citetads{2017MNRAS.464.4299F}, whose signal was probably enhanced by considering monthly-averaged data (as was the case for their analysis). 

As presented by different authors in the past, multiple chromospheric activity cycles are common among moderately-active stars (e.g., \citeads{1995ApJ...438..269B}, Ol\'ah et al. \citeyearads{2009A&A...501..703O}, \citeyearads{2016A&A...590A.133O}), and most of these can be associated with the \textit{active} and \textit{inactive} sequences proposed by \citetads{1998ApJ...498L..51B} and \citetads{2007ApJ...657..486B}. In the latter form, each sequence is characterized by a roughly constant number of rotation periods per activity cycle (i.e., the ratio $P_{\rm cyc}/P_{\rm rot}$). For example, \citetads{2015ApJ...812...12E} reports values of $n_{\rm short} \simeq 50$ and $n_{\rm long} \simeq 400$ (associated with the inactive and active branches, respectively), for the $\sim$1 Myr old solar analog HD~30495 (G1.5V, $P_{\rm rot} \simeq 11.36$ d, $P_{\rm long} \sim 12.2$~yr, $P_{\rm short} \sim 1.67$ yr). As discussed in Sect. \ref{sec_rotation}, we obtain a $P_{\rm rot} \sim 7.7$ d for $\iota$ Hor which leads to $n_{\rm short} = P_{2}/P_{\rm rot} \simeq 67$ and $n_{\rm long} = P_{1}/P_{\rm rot} \simeq 93$. Given the relatively small $n_{\rm long}$ value of $\iota$ Hor compared to HD~30495, it might look that only the short cycle ($P_{2}$) could be associated with the inactive sequence. However, our $n_{\rm long}$ value is roughly consistent with the active branch at the location of the star in the cycle period vs. rotation period diagram\footnote[11]{See \citetads{2016ApJ...826L...2M} for an updated version.} of \citetads{2007ApJ...657..486B}, as both sequences lie very close to each other in this region. This also explains why the ratio between the long and short cycle periods identified for $\iota$~Hor is smaller than for any other star with multiple periodicities (i.e., $P_{1}/P_{2} \sim 1.39$). For comparison, from the set of stars with complex cycles reported by \citetads{2016A&A...590A.133O}, the smallest\footnote[12]{We consider here the largest and smallest $P_{\rm cyc}$ values found by \citetads{2016A&A...590A.133O}.} value of $P_{\rm long}/P_{\rm short}$ is $1.84$ (HD131156B, K4V, $P_{\rm rot} \simeq 11.05$ d, $P_{\rm long} \sim 4.2$ yr, $P_{\rm short} \sim 2.28$ yr). Likewise from the compilation of stars with multiple cycle periods performed by \citetads{2017arXiv170409009B}, a minimum value of $P_{\rm long}/P_{\rm short} = 1.73$ is found (HD114710, F9V, $P_{\rm rot} \simeq 12.3$~d, $P_{\rm long} \sim 16.6$ yr, $P_{\rm short} \sim 9.6$ yr). The extreme location in the parameter space occupied by $\iota$ Hor motivates the importance of studying the evolution of its magnetic field, and how this field influences with the observed chromospheric and coronal activity. 

In connection with the previous discussion, we present below two phenomenological scenarios which could lead to the observed behavior in the chromospheric activity. For both cases, we consider as an additional constraint the information provided by the long-term X-ray monitoring of $\iota$~Hor (see \citeads{2013A&A...553L...6S}, \citeads{2016csss.confE.112S}, \citeads{2017AN....338..195S}). 

The first one takes elements from the solar case, assuming that the \textit{real} activity cycle is $\sim$$1.6$ yr as identified in the X-ray observations (given the larger cycle contrast in this wavelength range; see Sect. \ref{sect_1}). Our identified periods in the Ca~H\&K evolution could be then associated with a projection effect due to the $\sim$$60^{\circ}$ inclination of the star\footnote[13]{Estimated by \citetads{2010ApJ...723L.213M} assuming solid body rotation and previously reported values of $v\sin(i)$, $R_{*}$, and $P_{\rm rot}$.}, (less important for the optically-thin coronal emission), and a strong latitudinal migration of active regions. As has been shown in several Doppler-Imaging studies of main-sequence and pre main-sequence stars (see \citeads{2009A&ARv..17..251S}), magnetically-induced features (such as spots and faculae) seem to be concentrated towards the polar regions in very active stars, in contrast with the $\pm30^{\circ}$ latitudinal belts of solar activity. As $\iota$ Hor does not reach the very high activity levels of these objects, but is certainly more active than the Sun, an intermediate situation is expected. In this way, an asymmetry could be induced in the temporal evolution of the S-index values, due to an uneven distribution (in space and time) of chromospherically-active regions. More specifically, two signals with shorter and longer periods could be caused by the non-visibility, at certain times of the cycle\footnote[14]{Following the solar analogy, this would occur near activity maximum.}, of active regions from one hemisphere (located at latitudes larger than~$60^{\circ}$), combined with the opposite situation (i.e. contributions from both hemispheres) as the cycle progresses and the active regions move towards the equator. A similar geometrical effect was previously proposed by \citetads{2013A&A...553L...6S}, where the change in the cycle period would be related with asymmetries in the emerging activity between the visible and the partially visible hemispheres.

The second scenario considers that the two periods extracted from our analysis of the S-index variations, belong to \textit{independent activity cycles} (i.e. each one associated with a magnetic cycle). In this case, the X-ray emission should have contributions from \textit{both} magnetic cycles, and therefore, the observed period in the coronal emission corresponds to an \textit{apparent} mean value falling in-between the two real magnetic cycle periods. This would also imply that, with sufficient time coverage and sampling, both chromospheric cycle periods should appear in the X-ray evolution, with a similar beating pattern governing the variability of the coronal activity. While there might be hints of a secondary cycle in the X-ray data (see \citeads{2016csss.confE.112S}), the available observations are still more consistent with a single $1.6$ yr periodicity. Future monitoring of $\iota$ Hor in X-rays\footnote[15]{Additional observations with XMM-Newton are planned for the $2017-2018$ period (AO 16, PI: Sanz-Forcada).} may help to reveal if this second periodicity is present in the coronal variability of the star.

Finally, we would like to point out that these two scenarios should also imprint different signatures in the evolution of the magnetic field of the star. Assuming that the magnetic and activity time-scales are coupled as in the solar case, we expect a single polarity reversal over a period of 1.6 yr for the first-case scenario (i.e.~a 3.2 yr magnetic cycle). The situation is more complicated for the second case, as a double magnetic cycle would allow additional evolution patterns to occur (some of them could significantly deviate from the solar behavior). The ZDI reconstructions of the large-scale magnetic field of $\iota$ Hor, planned for the second paper of this study, could help us to establish if any of the previously discussed possibilities is occurring in the star. 

\subsection{$\boldsymbol\iota$ Horologii b}\label{sec_planet}

We now evaluate the robustness of our HARPSpol RV values by comparing them against the expected variation due to the presence of $\iota$ Hor b. While an independent orbital solution could be obtained from our data, we use instead the parameters derived by \citetads{2013A&A...552A..78Z}, as these have been determined from a much larger baseline and a consistent treatment of data from several instruments (see~Sect.~\ref{sect_2}). This also represents a more stringent approach, as we are not forcing a new best-fit solution but rather comparing our results directly with the expected behavior of the RV variations.

\begin{figure} 
\includegraphics[trim=0.4cm 0.8cm 0.0cm 0.3cm, clip=true, width=\linewidth]{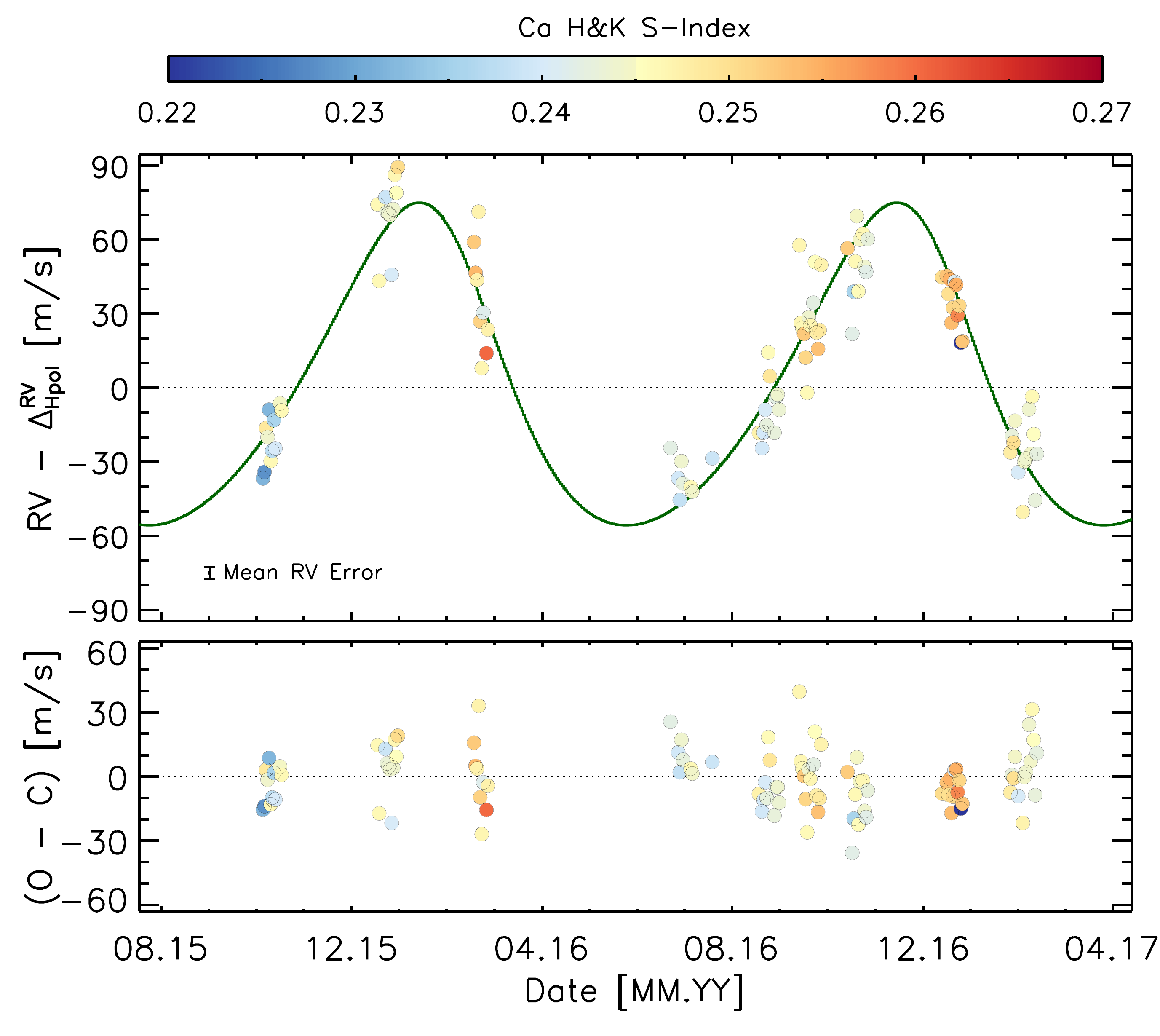}
\caption{Evolution of the RV variations of $\iota$ Hor determined from the HARPSpol observations. The top panel shows the RV variation relative to a fixed instrumental offset (see text for details). The solid line corresponds to the most recent orbital solution for $\iota$ Hor b (\citeads{2013A&A...552A..78Z}), and is used to compute the residuals in the bottom panel. Colors denote the associated Ca H\,\&\,K S-index value for a given observation (Sect. \ref{sec_activity}).}\label{fig_4}
\end{figure}

The \textit{Systemic 2} package (\citeads{2009PASP..121.1016M}) is used for this purpose, incorporating the HARPSpol RV values as an independent dataset and fixing all the orbital elements to the values provided by \citetads{2013A&A...552A..78Z}\footnote[16]{This includes a stellar mass for $\iota$ Hor of $M_{*} = 1.25$ M$_{\odot}$.}. Only two parameters are fitted in this process, namely the periastron passage time ($T_{0} = 2449738.24815$ d) and the RV reference level ($\Delta_{\rm Hpol}^{\rm RV} = 16923.38$ m s$^{-1}$), as it is necessary to adjust the phase (to include possible precession of the orbit with respect to our line-of-sight and the published solution) and the offset of the HARPSpol observations. The result shows very good agreement as illustrated in Fig. \ref{fig_4}, where the residuals show an rms value of $13.52$~m~s$^{-1}$ (slightly below the value reported by \citeads{2013A&A...552A..78Z}). This gives additional confidence to the RV values determined with the polarimetric mode of the HARPS spectrograph.       

One important aspect to consider is the possible influence due to the evolving activity (measured in terms of the $S_{\rm H}$ and $I_{\rm H\alpha}$ indicators) and the longitudinal magnetic field ($B_{\ell}$), over the RV variations of $\iota$ Hor. Following this idea, colors in Fig. \ref{fig_4} indicate the parallel evolution of the chromospheric activity of the star, given by the Ca H\,\&\,K S-index. This comparison does not reveal any evident relation between the activity, the amplitude of the RV variations, or the level of scatter of the residuals. However we note that, while the nine HARPSpol epochs provide relatively good phase coverage of the exoplanet orbit ($\sim$$1.6$ periods), their time-span falls short with respect to the activity cycle of the star (by $\sim$$30$ per cent), which displays epochs with considerably higher activity levels and variability (see Sect. \ref{sec_activity},~Fig.~\ref{fig_1}). 

\begin{figure*} 
\includegraphics[trim=0.3cm 0.6cm 0.0cm 0.2cm, clip=true, scale=0.228]{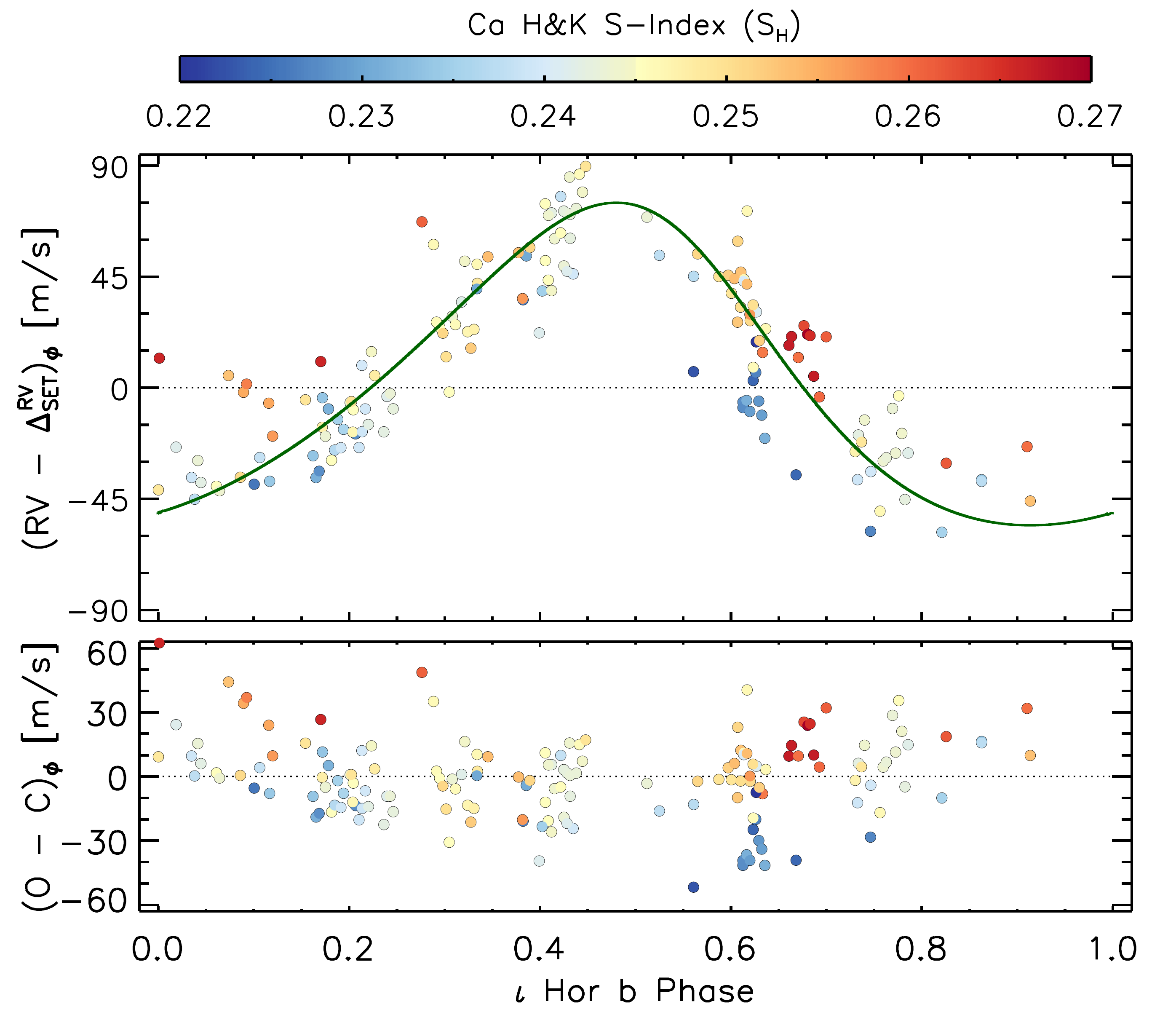}\includegraphics[trim=3.15cm 0.6cm 0.0cm 0.2cm, clip=true, scale=0.228]{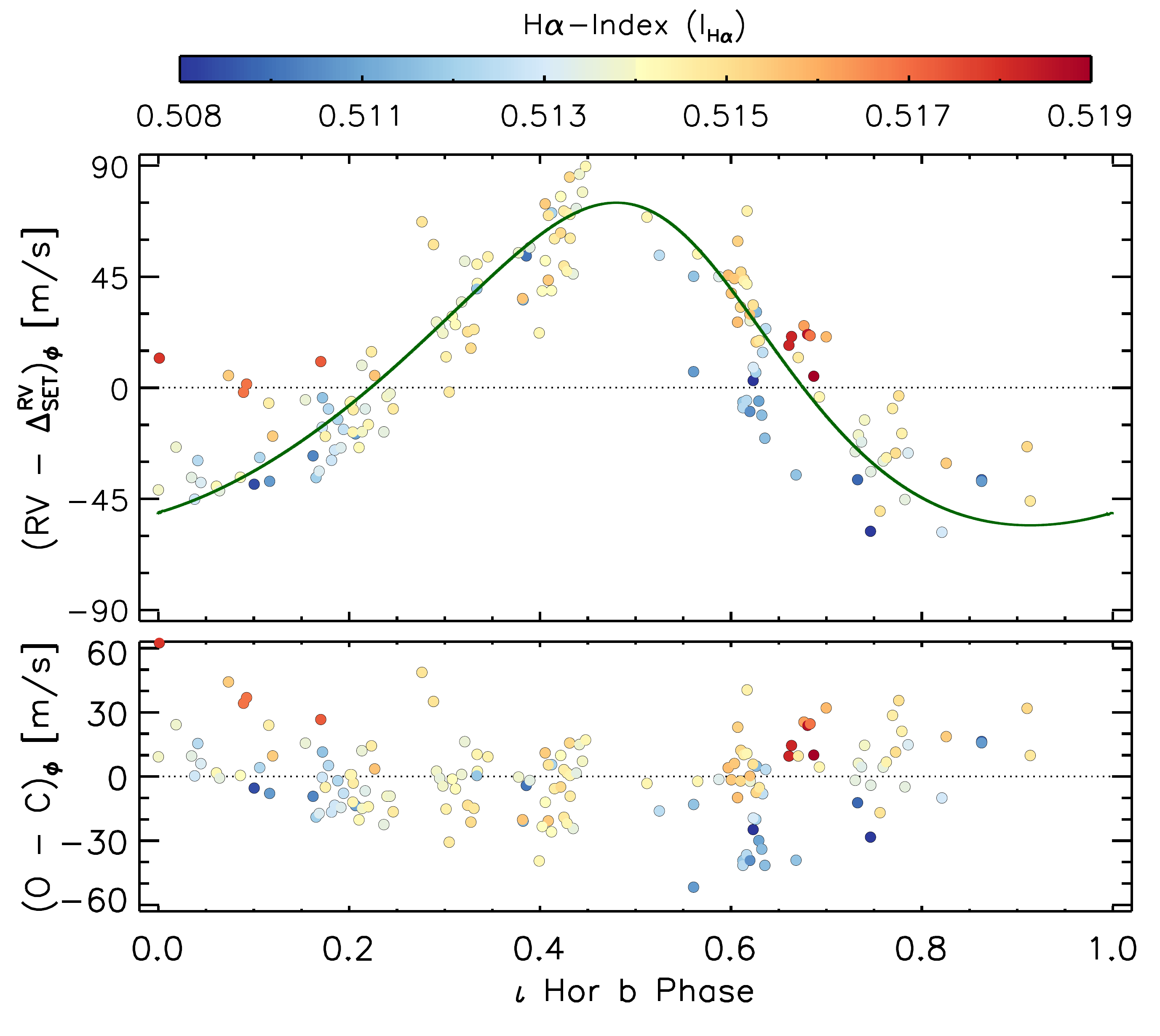}\includegraphics[trim=3.15cm 0.6cm 0.56cm 0.2cm, clip=true, scale=0.228]{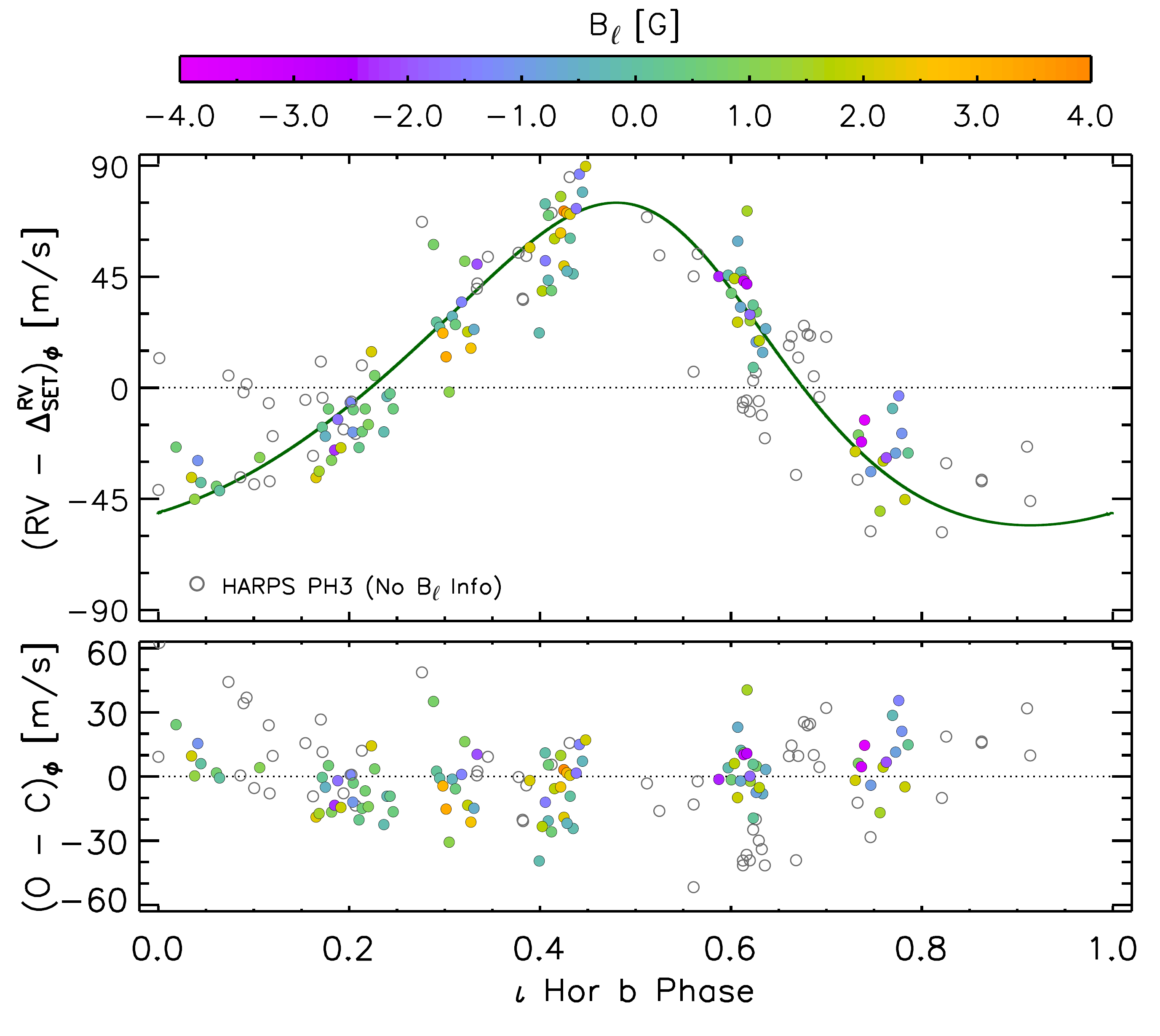}
\caption{RV variations of $\iota$ Hor determined from the combined HARPS PH3 and HARPSpol datasets. The phase and orbital solution (solid line) have been calculated using the parameters reported by \citetads{2013A&A...552A..78Z}. The color bars denote the corresponding values of $S_{\rm H}$ (left), $I_{\rm H\alpha}$ (middle), and $B_{\ell}$ (right) for each observation. The residuals are shown in the bottom panel.}\label{fig_5}
\end{figure*}

To improve this situation, we have expanded the baseline by including RV measurements from the HARPS PH3 observations (Sect. \ref{sec_archival}), extracted directly from the values reported by the pipeline\footnote[17]{\url{http://www.eso.org/rm/api/v1/public/releaseDescriptions/72}}. As we are considering a single spectrum per night, we have assumed conservative limits for the RV uncertainties, taking three times the value listed in the final data products (corresponding to photon noise alone), and adding a typical $2.0$ m s$^{-1}$ error in quadrature. The combined HARPS PH3 and HARPSpol RV results are presented in Fig. \ref{fig_5}, which have been now phase-folded to the orbital period of the exoplanet and include the evolution of the Ca~H\,\&\,K S-index (\textit{left}) and the H$\alpha$-index (\textit{middle}). The \textit{right} panel of Fig. \ref{fig_5} contains a similar plot for the corresponding surface-average longitudinal magnetic field values ($B_{\ell}$), only available for our HARPSpol observations.

There are some elements of Fig. \ref{fig_5} that are noteworthy. First of all, it is clear that the large variations of the RV of $\iota$ Hor are associated with a sub-stellar companion, and not with the ongoing activity-magnetic cycle. This can be seen from the fact that RV values consistent with the exoplanet orbit, have been observed at nearly all phases which display, at the same time, the full extent of activity levels in the star (Fig. \ref{fig_5}, \textit{left} and \textit{middle panels}). Likewise, as discussed in the Sun-as-a-star study performed by \citetads{2016MNRAS.457.3637H}, RV variations due to activity features show a strong correlation with the corresponding disc-averaged magnetic flux density. As this quantity should be proportional (to a first order) to the longitudinal magnetic field, the lack of coherence between the $B_{\ell}$ evolution and the RV signal (Fig. \ref{fig_5}, \textit{right}), further supports a non-magnetic origin for the observed long-term RV variations of $\iota$ Hor. 

On the other hand, an \textit{activity gradient} appears around the exoplanet orbit in the \textit{left} and \textit{right} panels of Fig. \ref{fig_5}. Assuming that the solution obtained by \citetads{2013A&A...552A..78Z} is a good representation of the system, this suggests that the activity level dominates the structure of the RV residuals. In this way, the RV measurements obtained during epochs of low activity are systematically lower compared with the expected values, whilst the opposite situation occurs for epochs of high activity. A similar pattern has been previously reported for the less active solar twin HIP 11915 (\citeads{2015A&A...581A..34B}), and considered as a signature of convective blue-shift suppression by localized magnetic fields. Determining how such behaviour could be modeled and removed from the observations is out of the scope of this paper. Nevertheless, such an investigation would clearly be of value for understanding stellar activity and its contribution as noise to RV planet searches. 


\subsection{Rotation period}\label{sec_rotation}

The last part of our analysis involves the short-term variability and the determination of the rotation period of the star ($P_{\rm rot}$). Literature values of this parameter range between $7.9$ d (\citeads{1997MNRAS.284..803S}) and $8.5$ d (\citeads{2010ApJ...723L.213M}). The former estimate was obtained from the level of chromospheric emission (expressed in the form of the $R^{\prime}_{\rm HK}$ indicator), and the empirical period-activity relation from \citetads{1984ApJ...279..763N}. The latter considered a LS periodicity analysis from a time-series of chromospheric Ca H\,\&\,K S-index values (between early-$2008$ and mid-$2010$ in Fig. \ref{fig_1}). In addition, similar $P_{\rm rot}$ values were inferred by \citetads{2011A&A...528A...4B}, using LS analysis over simulated activity signals in order to fit the short-term RV variations of the star (determined from short-cadence HARPS data).   

\begin{figure} 
\includegraphics[trim=0.4cm 20.1cm 0.0cm 0.4cm, clip=true, width=\linewidth]{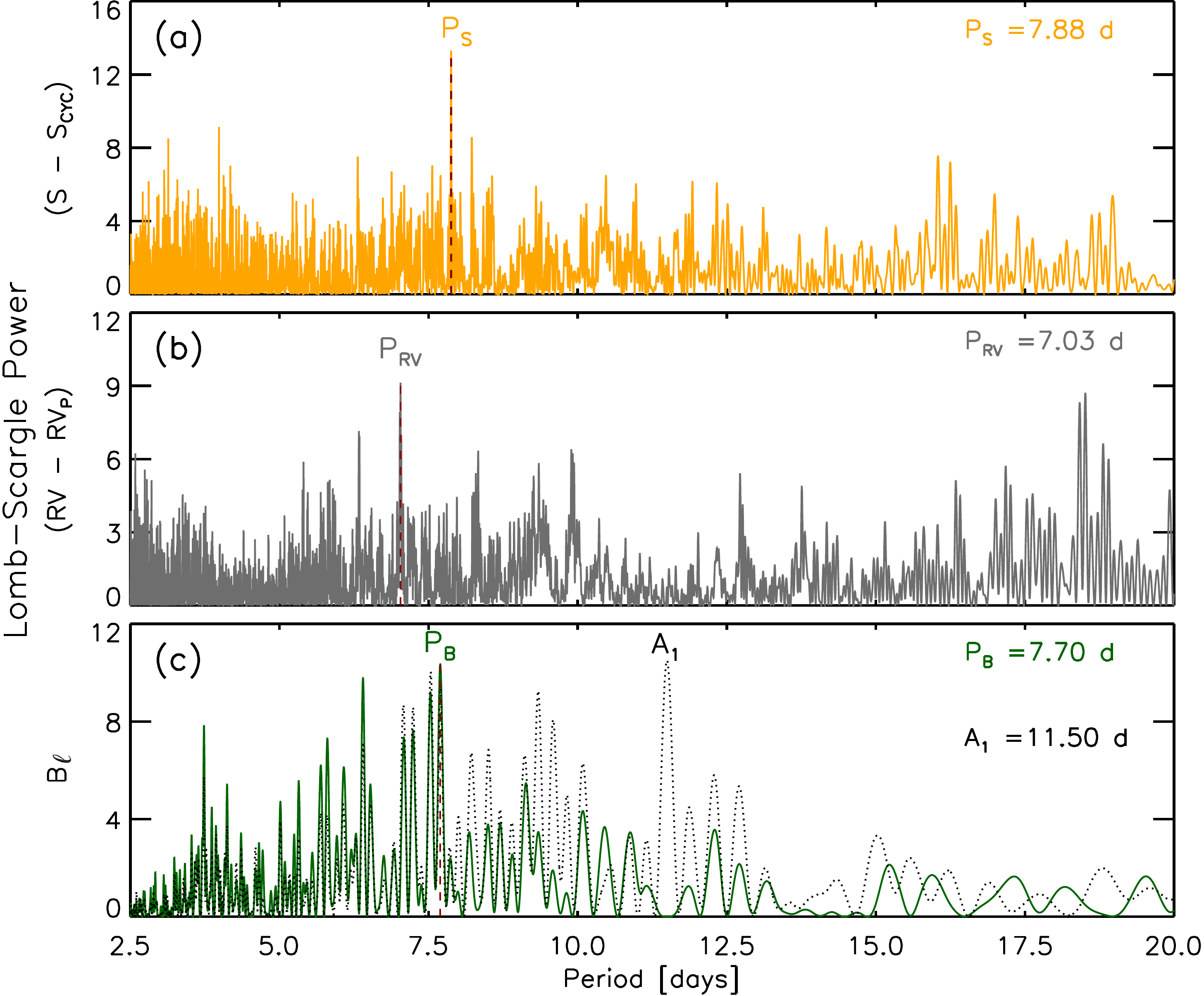}
\caption{Estimation of the rotation period ($P_{\rm rot}$) of $\iota$ Hor. Each panel contains the LS periodogram associated with the short-term variability of a specific parameter. After the removal of the dominant cycle ($S_{\rm cyc}$) and exoplanet (${\rm RV}_{\rm p}$) signals, the residual activity and RV variations are considered for panels \textit{a} and \textit{b}, respectively. The time-series of $B_{\ell}$ measurements listed in Table~\ref{table_2} are used to derive the raw (dotted line) and final (solid line) LS periodograms of panel \textit{c} (see text for more details). The identified period is indicated in each case.}\label{fig_6}
\end{figure}

Here we follow a combined approach, incorporating in the LS analysis the evolution of the Ca H\,\&\,K S-index, the RV variations, and the longitudinal magnetic field. As the activity cycle and the exoplanet signals are dominant in the former two cases, it is necessary to remove their contribution to extract any short-term periodicity from the data. This step has been carried out in order to produce the LS periodograms shown in Fig. \ref{fig_6} (panels \textit{a} and~\textit{b}), removing respectively our best fit 2-period activity cycle model ($S_{\rm CYC}$,~Sect.~\ref{sec_activity}), and the exoplanet RV signature (${\rm RV}_{\rm p}$,~Sect.~\ref{sec_planet}\footnote[18]{The removed signal was slightly different from the one presented in Sect. \ref{sec_planet}. The formal best fit orbital solution was used in this case ($P_{\rm orb} \simeq 308.3$ d, $M_{\rm p}\sin(i) \simeq 2.27$ M$_{\rm J}$, $e \simeq 0.16$, $\omega \simeq 64.3^{\circ}$), as required for a consistent extraction of the rotation period. The RV offsets (for HARPS and HARPSpol) and the $T_0$ value, were kept fixed with respect to the values used in Sect. \ref{sec_planet}.}). The panel \textit{c} in Fig. \ref{fig_6} shows the results derived from our series of $B_{\ell}$ measurements (Table \ref{table_2}). A raw LS periodogram of the data (dotted line) shows two prominent peaks located at $11.50$ d and $7.70$ d with nearly the same amplitude. However, as can be seen in panels \textit{a} and~\textit{b} of Fig.~\ref{fig_6}, the former peak does not appear in the LS periodograms of the chromospheric activity or the RVs. In addition, an $11.50$ d period would not only strongly disagree with the reports discussed in the previous paragraph, but also would be incompatible with the level of coronal emission observed in this star (see \citeads{2003A&A...397..147P}, \citeads{2013A&A...553L...6S}). Furthermore, this longer rotation period would be inconsistent with the allowed values from solid body rotation and published values of $R_{*}$ and $v\sin(i)$, including uncertainties (see Table \ref{tab_1})\footnote[19]{This means $\sin  i = (P_{\rm rot} \cdot v\sin i)/(2\pi R_*) > 1.0$.}. For these reasons, we interpret this period as a possible artifact which could be related to our data sampling (as most of our epochs are composed of 11 observations). The removal of this signal yields the final LS periodogram from this time-series (solid line in Fig.~\ref{fig_6}, panel \textit{c}).




\begin{figure} 
\includegraphics[trim=0.7cm 0.3cm 0.7cm 25.6cm, clip=true, width=\linewidth]{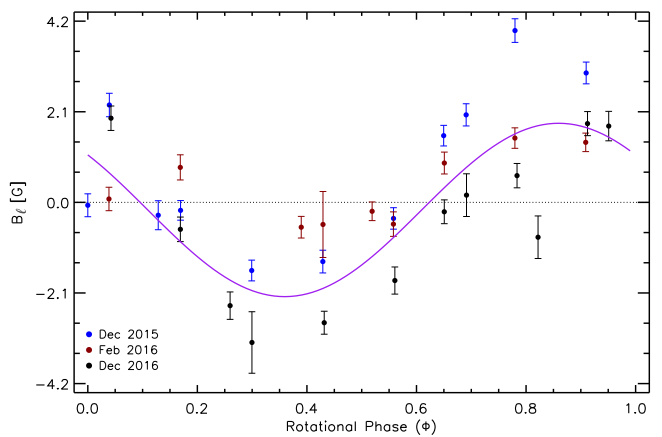}
\caption{Evolution of the longitudinal magnetic field of $\iota$ Hor as a function of rotational phase (with $P_{\rm rot} = 7.7$ d). Three separate epochs are presented, whose reference $\phi = 0$ dates have been adjusted individually. The solid line (not a fit) shows a phased $7.7$-d period sine curve as reference.}\label{fig_7}
\end{figure}

Given the low significance of the identified peaks in all cases, and the relatively clearer signal obtained from the $B_{\ell}$ dataset (even with much fewer data points), we take the $7.7$ d as the rotation period of the star and use the other two values as conservative limits for the uncertainty. Additionally, as illustrated in Fig. \ref{fig_7}, this value appears directly by looking at the evolution of $B_{\ell}$ in certain epochs, which is not the case for the any of the activity proxies or the RV variations. This corroborates the findings of \citetads{2016MNRAS.461.1465H}, which showed that time-series of $B_{\ell}$ measurements provide a more robust signature for the determination of $P_{\rm rot}$ compared to other activity indicators or the RVs. 

\section{Summary}\label{sect_6}

We presented here the initial results of our long-term monitoring campaign of $\iota$ Hor using the spectropolarimetric capabilities of the HARPS spectrograph. This paper explored the different time-scales of variability on the magnetic activity and radial velocity of the star, and provided information concerning the observing strategy and the data reduction process. Measurements of the surface-averaged longitudinal magnetic field of this object are also reported here for the first time. The observed range of variability for this parameter ($\pm4$ G) is fully consistent with the scatter reported in snapshot observations of stars of similar spectral type and age (\citeads{2014MNRAS.444.3517M}).          

Our analysis revisited the activity cycle of $\iota$ Hor, showing that the long-term evolution of the chromospheric activity can be fairly reproduced by a beating pattern, resulting from the superposition of two, out-of-phase, periodic signals of similar amplitude (Sect. \ref{sec_cycles}). The identified periods and their corresponding amplitudes are: $P_{1} \simeq 1.97 \pm 0.02$~yr with $A_{1} = 0.0118 \pm 0.0010$ and $P_{2} \simeq 1.41 \pm 0.01$~yr with $A_{2} = 0.0102 \pm 0.0008$. Two different physical scenarios were proposed and discussed in order to place this particular behavior in the general solar-stellar context.  

We also presented here $101$ new radial velocity measurements of $\iota$ Hor, which showed a remarkably good agreement with the expected evolution due to the presence of the exoplanet of this system (Sect. \ref{sec_planet}). By simultaneously tracing the evolution of the magnetic activity and the radial velocity, we find evidence of the former controlling the structure of the residual variations in the latter (once the signal from the exoplanet has been removed). While this poses a challenge in the search for additional planets in this system, characterising this behaviour in a larger sample of active planet-hosting stars could represent a step forward in our understanding of stellar activity in the exoplanet context.


Finally, we estimated here a rotation period for $\iota$ Hor of $\sim$$7.7$ d, roughly in agreement with the lower end of previous reports in the literature (Sect. \ref{sec_rotation}). This value is likely to be improved using the information contained in the shapes of the LSD Stokes V profiles, and the optimization routines applied during the Zeeman Doppler Imaging analysis planned for a future study.


\section*{Acknowledgements}

We would like to thank the referee, Martin K\"urster, for his constructive comments which helped to improve the quality of this paper. J.D.A.G. was supported by \textit{Chandra} grants AR4-15000X and GO5-16021X. J.J.D. was supported by NASA contract NAS8-03060 to the \textit{Chandra} X-ray Center. J.S.F. acknowledges support from the Spanish MINECO through grant AYA2014-54348-C3-2-R. Support for O.C. and S.P.M. is provided by NASA Astrobiology Institute grant NNX15AE05G. C.G. was supported by \textit{HST} GO-13754 and \textit{Chandra} GO7-18017X grants. Based on observations collected at the European Organisation for Astronomical Research in the Southern Hemisphere under ESO programmes 096.D-0257, 097.D-0420 and 098.D-0187. We are thankful to the different ESO OPC reviewers for the allocation of the HARPSpol observing time in each semester, and with the ESO OPO for accommodating the required observing strategy for this project. J.D.A.G. would like to thank the staff of the ESO La Silla Observatory for the strong support during each observing run of this programme. Based on data obtained from the ESO Science Archive Facility under request number jalvarad.212739. This work has made use of the VALD database, operated at Uppsala University, the Institute of Astronomy RAS in Moscow, and the University of Vienna. This research has made use of the SIMBAD database, operated at CDS, Strasbourg, France. This research has made use of NASA's Astrophysics Data System. 




\bibliographystyle{mnras}
\bibliography{Biblio} 






\bsp	
\label{lastpage}
\end{document}